\documentclass[usenatbib]{mn2e}
\usepackage{graphicx,times}
\usepackage{bm}
\usepackage{epsf}

\newcommand{\rX}{\mathrm X}
\newcommand{\rY}{\mathrm Y}
\newcommand{\rZ}{\mathrm Z}

\newcommand{\on}{{\hat \Omega}}
\newcommand{\beq}{\begin{equation}}
\newcommand{\eeq}{\end{equation}}
\newcommand{\beqa}{\begin{eqnarray}}
\newcommand{\eeqa}{\end{eqnarray}}
\newcommand{\n}{\noindent}
\newcommand{\oh}{\hat \Omega}
\def\qtwo{\qquad\qquad}

\def\hhs{\delta\delta\psi}

\def\myc{{\cal C}}
\def\myC{{\cal C}}

\date{\today,~ $ $Revision: 0.9 $ $}

\begin{document}

\onecolumn

\title[Secondary non-Gaussianity and Cross-Correlation Analysis]
{Secondary non-Gaussianity and Cross-Correlation Analysis}

\author[Munshi, Valageas, Cooray \& Heavens]
{Dipak Munshi$^{1,2}$, Patrick Valageas$^{3}$, Asantha Cooray$^{4}$ and Alan Heavens$^{1}$ \\
$^{1}$Scottish Universities Physics Alliance (SUPA),~ Institute for Astronomy, University of Edinburgh, Blackford Hill,  Edinburgh EH9 3HJ, UK \\
$^{2}$School of Physics and Astronomy, Cardiff University, CF24 3AA \\
$^{3}$Institut de Physique Th\'eorique, CEA Saclay 91191,Gif-sur-Yvette, France\\
$^{4}$ Department of Physics and Astronomy, University of California, Irvine, CA 92697}
\maketitle 
\begin{abstract}
We develop optimised estimators of two sorts of power spectra for fields defined on the sky, in the presence of partial sky coverage.  The first is the cross-power spectrum of two fields on the sky;  the second is the skew spectrum of three fields.  The cross-power spectrum of the Cosmic Microwave Background (CMB) sky with tracers of large-scale-structure is useful as it provides valuable information on cosmological parameters.  Numerous recent studies have proved
the usefulness of cross-correlating CMB sky with external data sets, which
probes the Integrated Sachs Wolfe Effect (ISW) at large angular scales and the Sunyaev
Z\'eldovich (SZ) effect from hot gas in clusters at small angular scales.  The skew spectrum, recently introduced by Munshi \& Heavens (2009), is an optimised statistic which can be tuned to study a particular form of non-gaussianity, such as may arise in the early Universe, but which retains information on the nature of non-gaussianity.  In this paper we develop the mathematical formalism for the skew spectrum of 3 different fields.  When applied to the CMB, this allows us to explore the contamination of the skew spectrum by secondary sources of CMB fluctuations.   
Considering the three-point function, the study of the bispectrum provides valuable
information regarding  cross-correlation of secondaries with lensing of CMB with much higher significance compared to just the study 
involving CMB sky alone. After developing the analytical model we use them to
study specific cases of cosmological interest which include cross-correlating
CMB with various large scale tracers to probe ISW and SZ effects for
cross spectral analysis. Next we use the formalism to study the signal-to-noise
ratio for detection of the weak lensing of the CMB by cross-correlating it with
different tracers as well as point sources for CMB experiments
such as Planck. 
\end{abstract}
\section{Introduction}
Observations of cosmic microwave background (CMB) and that of large scale structure carry 
complementary cosmological information. While all-sky CMB observations such as NASA's WMAP \footnote {http://map.gsfc.nasa.gov/}
and ESA's current Planck \footnote {http://www.rssd.esa.int/index.php?project=Planck} experiments primarily  probes the distribution of matter and radiation 
at redshift $z=1300$, large-scale surveys tend to give us a window at lower redshift $z \sim 0$.
The main advantage of cross-correlating such independent data sets lies in the fact that it is
possible to highlight signals which may not be otherwise detected in individual data
sets independently. Earlier studies in this direction include, \citet{Peir} 
who carried out 
a detailed error forecast of such cross-correlation analysis for cosmological parameters.
Clearly, for these tracers to be effective in constraining cosmology, they should be as numerous as
possible to reduce the Poisson noise and the survey should cover as large a fraction of
the sky as possible to reduce sample variance.

Various authors have used different external data sets with specific astrophysical
tracers to trace the large-scale structure (LSS), with one of the main motivations being to detect the ISW effect as
predicted for $\Lambda$CDM cosmology. Earlier studies in this direction include 
\cite{FosGaz1,FosGaz2}
who cross-correlated the SDSS Data Release 1 galaxies with the first-year full sky WMAP data.
Nolta et al. (2003) cross-correlated the NVSS radio source catalogue with first-year full sky WMAP data.
Scranton et al. (2003) who correlated Sloan Digital Survey against WMAP data. 
\citet{BoughCrit1,BoughCrit2,BoughCrit3} 
used two tracers of the large-scale structure: the HEAO1 A2 full sky hard X-ray map and NVSS 
full sky radio galaxy survey. A maximum likelihood 
fit to both data sets yields a detection of an ISW amplitude at a level consistent with what is
predicted by the $\Lambda$ CDM cosmology. Most of these studies detected ISW effect at a level 
of 2-3 $\sigma$ although the error analysis models and the statistic used were sometimes completely 
different (see \citet{Ho08} for tomographic studies involving ISW and \citet{Hi08} for weak lensing detection).
The ISW effect remains one of the most direct and quantitative
measure of the dark energy available to us today. Future all-sky missions such as Planck will 
provide an excellent possibility to extend these studies to higher confidence regime.
While the above studies are mainly focussed on large angular scales, where the ISW effect plays an important role,
at small angular scales, the presence of clusters and probably the associated filamentary network in which they reside 
can also affect both CMB maps through the Sunyaev-Zeldovich effect \citep{SZ} as well as through the X-ray maps 
via bremsstrahlung. Cross-correlation analysis of the diffuse soft X-ray background maps of 
ROSAT with  WMAP 1st year data were performed by \citep{Die1,Die2}. This study was 
motivated by the fact that hot gas in clusters can be more easily detected by cross-correlating 
X-ray and CMB maps. Although no evidence was found of this effect it opens the possibility of detecting 
such an effect in future high-resolution CMB maps. 
All these act as a motivation for development of a generic techniques to cross-correlate high-resolution 
CMB maps with other maps from LSS surveys.   In this paper we focus on cross-correlating two or three different datasets, but the challenges are similar to those arising from a single dataset.  For example, the
estimation of the power spectrum from a single high-resolution 
map poses a formidable numerical problem in terms of computational requirements. 
Typically two different methods are followed.  The first os the non-linear maximum likelihood method,
or its quadratic variant, which can be applied to smoothed degraded maps, as it is not 
possible to directly invert a full pixel covariance matrix \citep{teg}. 
To circumvent this problem a pseudo-${\cal C}_{\ell}$s (PCL) technique was invented 
\citep{Hiv} 
which is unbiased though remains suboptimal. In recent analysis \citet{Efs1,Efs2} has shown how
to optimize these estimators which can then be used to analyse high-resolution maps in a very fast 
and accurate way. We generalize the PCL-based approach here to compute the cross-correlation of different data sets. The method developed here is completely 
general and can be applied to an arbitrary number of data sets. For example, our formalism can analyse the degree 
of cross-correlation among various CMB surveys observing the same region of the sky with
different noise levels and survey strategies.

%
%
For near-Gaussian fields, two-point analysis from any cosmological survey provides the bulk of the cosmological information. 
Nevertheless, going one step further,
at the level of three-point correlation, the detection of departure from Gaussianity in the CMB
can probe both primary non-Gaussianity see (e.g. \cite{MuHe09})
as well as the mode-coupling effects due to secondaries.  The possibility of further improving a detection
of primordial non-Gaussianity with CMB maps, given current hints with WMAP data \citep{YaWa08,SmSeZa09},
provides further motivation in this direction. One of the prominent contributions to the secondary non-Gaussianity
is the coupling of weak lensing and sources of secondary contributions such as SZ \citep{GoldbergSpergel99,CoorayHu}. 
Although weak lensing  produces a characteristic signature  in the CMB angular power spectrum, 
its detection has proved to be difficult internally from CMB power spectrum alone.
The non-Gaussianity imprinted by lensing into the primordial CMB remains below the 
detection level of current experiments, although with Planck the situation is likely
to improve. The difficulty originates mainly due to the fact that such detections are linked to
the four-point statistics of the lensing potential. However cross-correlating CMB data with 
external tracers means lensing signals can be probed at the level of the mixed bispectrum. 
After the first unsuccessful attempt
to cross-correlate WMAP against SDSS, recent efforts by \cite{SmZaDo00} have found a clear 
signal of weak lensing of the CMB, by cross-correlating WMAP against NVSS which covers a significant 
fraction of the sky. Their work also underlines the link between three-point statistics estimators 
and the estimators for weak lensing effects on CMB.

The study of non-Gaussianity is primarily focused on the bispectrum \citep{Heav98}, however in practice
it is difficult to probe the entire configuration dependence in the harmonic space from noisy data.
The cumulant correlators are multi-point correlators collapsed to probe two-point statistic. 
These were introduced in the context of analyzing galaxy clustering by \citet{Szapudi}, and were later found to be 
useful for analyzing projected surveys 
such as APM \citep{Mun}. Being two-point statistics they can be analyzed in the multipole space by defining an associated 
power-spectrum.  Recent studies by \cite{Cooray3} and \citet{Cooray8} have shown its wider applicability including e.g. in 
21cm studies. However, the multi-spectrum elements defined in multipole space are difficult to estimate directly 
from the data because of their complicated response to partial sky coverage and inhomogeneous 
noise, as well as associated high redundancy in the information content.
However such issues are well understood in the context of power-spectrum analysis.
Borrowing from previous results, in this paper we show how the cross-power spectrum and the skew spectrum can be studied in real data in an optimal way. We concentrate
on two effects: the cross-correlation power spectrum, which is recovered by cross-correlating two different 
(but possibly correlated) data sets, focussing on weak lensing effects on the CMB, and secondly the contributions to the skew spectrum from foreground effects.
The relation of such cross-power spectrum estimators with higher-order multi-spectra such as the bispectrum is also 
discussed in the context of methods known as pseudo-$C_l$s and quadratic estimators. 
We derive the error-covariance matrices and discuss their validity  in the signal- 
and noise-dominated regimes and comment on their relationship to the Fisher matrix.

This layout of the paper is as follows: in \textsection2 we use the formalism based on Pseudo-${\cal C}_{\ell}$ analysis
for power-spectra to study the cross-correlation power spectrum of different data sets. 
While we keep the analysis completely
general, it is specialized for the case of near all-sky analysis and use it to compute the signal-to-noise 
and the covariance of estimated ${\cal C}_l$s  for various tracers with Planck-type all-sky experiments.
Possibilities of  using various weights which can make the pseudo-${\cal C}_{\ell}$ approach near optimal
in limiting cases of the signal-dominated regime or the noise-dominated regime are also discussed.
In \textsection3 we continue our discussion on Pseudo-${\cal C}_{\ell}$ but generalize it to the
analysis of the skew spectrum. Such an estimator can handle the partial sky
coverage and noise in a very straightforward way. However in general it remains sub-optimal. 
In the high-$l$ regime where mode-mode coupling can be modelled using a fraction of sky $f_{sky}$
proxy one can make such an estimator nearly optimum using a suitable weighting. After a very brief introduction
to various physical effects in \textsection4 which introduce mode-mode coupling that leads to CMB bispectra, we move
on to develop a crude but fast estimator for the skew spectrum in 
\textsection5.   Section \textsection6 is devoted to developing 
the mixed bispectrum analysis in an optimal way by introducing inverse covariance weighting
of the data. We analyze both one-point and two-point
collapsed bispectral analysis. The one-point estimator or the mixed skewness is introduced - being a 
one point estimator it compresses all available information in a bispectrum to a single number.
Next, we introduce the mixed  skew spectrum which compresses various
components of a bispectrum to a power spectrum in an optimum way. Next, in \textsection7, the general
formalism of bispectral analysis is used for specific cases of interest.
\section{The Pseudo-${\cal C}_{\ell}$ Estimator for Cross-correlation Analysis}
\label{sec:intro}

In this section we generalise results from pseudo-${\cal {C}}_\ell$ power spectrum estimation of a single field with partial sky coverage, to cross-power spectra of two fields.
For two fields $\Phi^X$, $\Phi^Y$ defined on the sky, the pseudo-${\cal C}_{\ell}$ estimators are constructed from the spherical 
harmonic transforms $\tilde a_{lm}^{X,Y}$  over the partial sky, where the fields are assumed to take zero value in unobserved regions.

The $\tilde a_{lm}^{X,Y}$  are related to the true all-sky spherical harmonics 
$a^{X,Y}_{\ell m}$ by a linear transformation,
via the transformation matrix $K_{lm l'm'}$, and where possible direct inversion to get $a_{lm}$ 
is much faster than maximum-likelihood analysis. Using a suitable
choice of weighting function for the data this estimation can also be made nearly optimal.

\begin{figure}
\begin{center}
{\epsfxsize=10. cm \epsfysize=5. cm {\epsfbox[23 439 590 715]{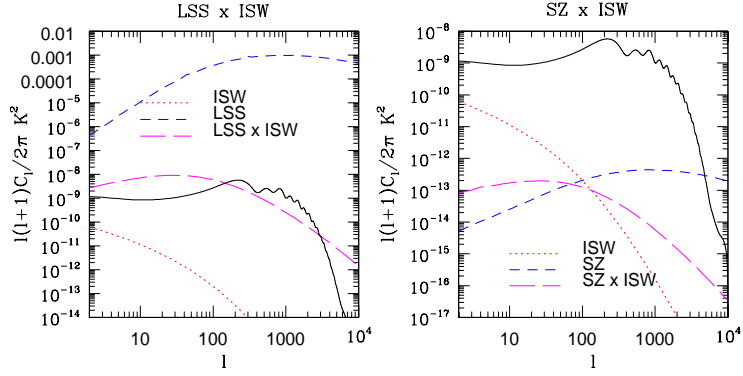}}}
\end{center}
\caption{The power-spectra $C_l$s are plotted for various analysis. In the left panel
we show the power spectrum corresponding to a LSS tracer such as NVSS  and that of CMB ISW.
The power spectra associated with the two is also depicted. In the right panel we show
SZ and ISW power-spectrum and their cross-correlation. The CMB power spectrum is also
plotted in both panels. SZ curves include only the part that is correlated with the large-scale density field.}
\label{fig:bicls1}
\end{figure}

\subsection{Estimator for $\tilde {\cal C}_l ^{X,Y}$}

\n
The transformation from the pixel-space to harmonic domain can be expressed as follows:

\begin{equation}
\tilde a_{lm}^{X,Y} = \sum_{pixels i} \Phi^{X,Y}(\hat\Omega_i) w(\hat\Omega_i)^{X,Y} \Omega_p(\hat \Omega_i)^{X,Y} Y_{lm}(\hat \Omega_i) = \sum_{l'm'}a_{l'm'}^{X,Y}
K_{lml'm' }^{X,Y}.
\end{equation}

\n
Here $w(\hat\Omega_i)$ denotes the pixel space weight, $Y_{lm}(\hat \Omega_i)$ represents the spherical harmonic
and $\Omega_p(\hat \Omega_i)$ is the pixel area (which we will assume independent of the pixel position).
Expanding the weighting function in a spherical basis one can write down the coupling matrix $K_{l_1m_1l_2m_2}$
which encodes all information regarding mode-mode coupling due to partial sky coverage as (see e.g. \citet{Hiv} for detailed derivation):
 
\begin{eqnarray}
K_{l_1m_1l_2m_2} &=& \int w({\on}) Y_{l_1m_1}(\on) Y_{l_2m_2}(\on) {\rm d} \on \nonumber \\
&=& \sum_{l_3m_3} \tilde w_{l_3m_3} \left( {(2l_1+1)(2l_2+1)(2l_3+1)\over 4\pi } \right)^{1/2} 
\left ( \begin{array}{ c c c }
     l_1 & l_2 & l_3 \\
     0 & 0 & 0
  \end{array} \right)
\left ( \begin{array}{ c c c }
     l_1 & l_2 & l_3 \\
     m_1 & m_2 & m_3
  \end{array} \right),
\end{eqnarray}

\noindent
where $\tilde w_{lm}$ is the transform of the window or (arbitrary) weighting function.
The matrices represents $3J$ functions. The quantum numbers $l$ and $m$ need to satisfy certain conditions 
for the $3J$ functions to have non-vanishing values.
The pixel area of maps $\Phi^X$  and $\Phi^Y$ will be denoted by 
and the associated
weights with each pixels will be left arbitrary $w_i^{X,Y}$. Adopting the notation of \citet{Efs1}, we write the pseudo-${\cal C}_{\ell}^{X,Y,XY}$ in terms of the underlying 
true power-spectrum ${\cal C}^{X,Y,XY}_{\ell}$s.

Defining the following:
\begin{equation}
\langle \tilde {\cal C}^{X}_{\ell} \rangle =  { 1 \over (2\ell+1) } \sum_m | {\tilde a^{X}}_{lm} |^2;
~~~ \langle \tilde {\cal C}^{XY}_{\ell} \rangle =  { 1 \over (2\ell+1) } \sum_m {\mathrm Real} ({\tilde a}^{X}_{lm}*{\tilde a}^{Y}_{lm})
\end{equation}
and similarly for $\langle \tilde {\cal C}^{X}_{\ell} \rangle$, we have
\begin{equation}
\langle \tilde {\cal C}_{\ell}^{\alpha}  \rangle = \sum_{\ell} C_{\ell'}^{\alpha} M_{\ell\ell'}^{\alpha},
\end{equation}
where $\alpha=X,Y$ or $XY$, and we can estimate the true covariance matrices as
\begin{equation}
\hat C_{\ell}^{\alpha} = (M^{-1})^{\alpha}_{ll'} {\tilde C}^{\alpha}_{\ell'}.
\end{equation}

\noindent
where the  matrix $M$ can be expressed in terms of $3J$ symbols as \citep{Hiv}:

\begin{equation}
M_{{\ell_1}{\ell_2}}^{\alpha} =  (2\ell_2+1) \sum_{\ell_3} {(2\ell_3 + 1) \over 4 \pi} \tilde W_{l_3}^{\alpha}
{\left ( \begin{array}{ccc}
        \ell_3 & \ell_2 & \ell_1  \\
        0  & 0 & 0
       \end{array} \right )^2}.
\end{equation}

\n
and $W_l= {1 \over 2l+1}\sum_m |w_{lm}|^2 $ is the power spectra associated
with the mask. Note that the transformation matrices $M^{X,Y}$ depends on the power-spectrum of the weighting function 
$w_{lm}^{X,Y}$, whereas the matrix $M^{XY}$ for cross-power spectra is determined by
the cross-power spectra $\tilde W_{l_3}^{XY}$ of two weighting functions.  Independent of the  
choice of weights the estimators $\cal C_{\ell}^\alpha$  remain unbiased.

\subsection{Covariances of Pseudo-${\cal C}_{\ell}$s}
\label{sec:pseudo}

\n
The pseudo-${\cal C}_{\ell}$s are unbiased. The variances of these estimators can be computed
analytically for arbitrary sky coverage and a non-uniform Gaussian noise distribution. The
deviation of the estimated $\tilde C_l$ from the ensemble average $\langle \tilde C_l \rangle$ 
is denoted by $\delta \tilde C_l$.

\begin{equation}
\delta \tilde {\cal C}_l^\alpha = \tilde {\cal C}_l^\alpha - \langle \tilde {\cal C}_l^\alpha \rangle.;
\qtwo \alpha~ \epsilon~ X,Y,XY
\end{equation}

\n
We are concerned here with the computation of the covariance of estimated $C_ls$. We begin by defining 
the covariance matrix:

\begin{equation}
\langle \delta \tilde {\cal C}_l^\alpha \delta \tilde {\cal C}_{l'}^\beta \rangle = \langle \tilde {\cal C}_l^\alpha
\tilde {\cal C}_{l'}^\beta \rangle 
- \langle \tilde {\cal C}_l^\alpha \rangle \langle \tilde {\cal C}_{l'}^\beta \rangle; \qtwo \alpha,\beta ~\epsilon~ X,Y,XY
\end{equation}

\n
The covariance of ${\cal C}_ls$ from individual surveys $\Phi^{X}$ or $\Phi^{Y}$
can be expressed as follows \citep{Efs1}.  

\begin{eqnarray}
\langle \delta \tilde {\cal C_{\ell}}^X \delta \tilde {\cal C_{\ell'}}^Y \rangle =
\sum_L \Big \{ {\cal C_{\ell}}^X{\cal C_{\ell}}^X   { 1 \over 2L +1}
\sum_M | w_{LM}^X |^2 |  w_{LM}^X |^2 + {\cal C_{\ell}}^X
{ 1 \over 2L +1} \sum_M | w_{LM}^X |^2 |  (w \sigma^2)^Y_{LM} |^2 \nonumber \\
+ { 1 \over 2L +1} \sum_M | w_{LM}^X |^2 |  (w \sigma^2)^Y_{LM} |^2 \Big \}
{\left ( \begin{array}{ccc}
        L & \ell & \ell'  \\
        0  & 0 & 0
       \end{array} \right )^2}.
\end{eqnarray}

\noindent
Extending the above results similarly the covariance of ${\cal C}^{\chi}_l$ for the cross-power-spectrum can
be expressed as:

\begin{eqnarray}
\langle \delta {\tilde \myC_{\ell}}^{XY} \delta {\tilde \myC_{\ell'}}^{XY} \rangle = &&
\sum_L \Big \{  {\cal C_{\ell}}^X{\cal C_{\ell}}^Y   { 1 \over 2L +1}
\sum_M | w_{LM}^X |^2 | w_{LM}^Y |^2
+   {\cal C_{\ell}}^X { 1 \over 2L +1}
\sum_M | w_{LM}^X |^2 | (w^2 \sigma^2)_{LM}^X |^2 \noindent \\
&& +   {\cal C_{\ell}}^Y { 1 \over 2L +1}
\sum_M | w_{LM}^X |^2 | (w^2\sigma^2)_{LM}^Y |^2 
+   { 1 \over 2L +1}
\sum_M |  (w^2\sigma^2)_{LM}^X |^2 | (w^2 \sigma^2)_{LM}^Y |^2 \Big \}
{\left ( \begin{array}{ccc}
        L & \ell & \ell'  \\
        0  & 0 & 0
       \end{array} \right )^2},
\end{eqnarray}

\n
In our derivation we have assumed that all three power spectra are being
estimated from the data simultaneously.

\begin{figure}
\begin{center}
{\epsfxsize=10. cm \epsfysize=5. cm {\epsfbox[23 439 590 715]{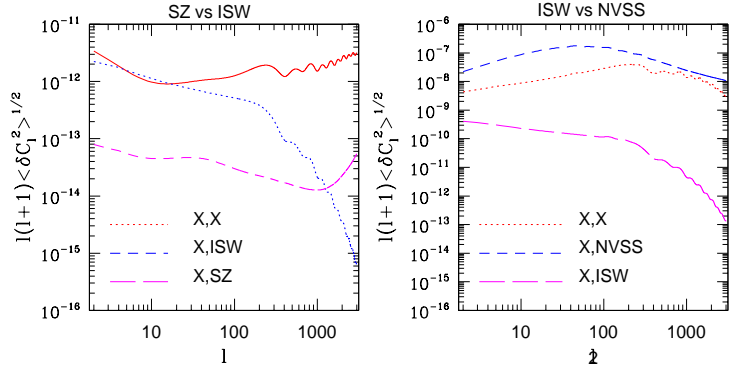}}}
\end{center}
\caption{The variance for estimated $C_l$s are plotted. The variance for estimated $C_l$ corresponding to 
ISW, SZ and their cross-correlation are plotted in the left panel. The ISW vs local tracers (NVSS type)
analysis is plotted in the right. The number density of galaxies for NVSS-type survey was taken to be
$\bar N = 7 \times 10^8$ per Steradian. For CMB a Planck type experiment was assumed. Results are for
all-sky surveys. Results plotted are for $f_{sky}=1$. For near all sky survey, the variances will 
scale linearly with $f_{sky}$.}
\label{fig:bicls2}
\end{figure}

\noindent
The three off-diagonal terms can be similarly expressed as:
\begin{equation}
\langle \delta {\tilde \myC_{\ell}}^X \delta {\tilde \myC_{\ell'}}^Y \rangle = \sum_L ({\cal C_{\ell}}^X)^2 { 1 \over 2L +1}
| w_{LM}^X w_{LM}^{Y*} |^2 { \left( {2L+1 \over 2 \pi}  \right) }
{\left ( \begin{array}{ccc}
        L & \ell & \ell'  \\
        0  & 0 & 0
       \end{array} \right )^2, } 
\end{equation}

\begin{eqnarray}
\langle \delta {{\tilde \myC}^{XY}_{\ell}} \delta {{\tilde \myC}^Y_{\ell'}} \rangle = 
\sum_L {\cal C_{\ell}}^{XY}{\cal C_{\ell}}^X \Big \{  { 1 \over 2L +1} 
\sum_M | w_{LM}^Xw_{LM}^{Y*} | |  w_{LM}^X |^2 + {\cal C_{\ell}}^{XY}{ 1 \over 2L +1} 
\sum_M  | (w^2 \sigma^2)_{LM}^X w_{LM}^{Y*} | \Big \}
{ \left( {2L+1 \over 2 \pi}  \right) }
{\left ( \begin{array}{ccc}
        L & \ell & \ell'  \\
        0  & 0 & 0
       \end{array} \right )^2}.
\end{eqnarray}

\noindent
Here we have introduced following notations:

\begin{eqnarray}
&& w^X_{lm} = \int d\Omega\, w^X(\hat \Omega) Y_{lm}(\hat \Omega) \\
&& (w\sigma^2)_{lm}^X = \int d\hat \Omega w^X(\hat \Omega) \sigma_X^2(\Omega) Y_{lm}(\hat \Omega)  \\
&& (w^2\sigma^2)^X_{lm} = \int d \hat \Omega {w^X}(\hat \Omega)^2 \sigma_X^2(\hat \Omega) Y_{lm}(\hat \Omega).
\end{eqnarray}
  
\noindent
Similar expression hold for the second survey and the cross terms for product of two surveys are also 
needed to derive the error covariance matrices. Finally the error covariances associated with 
deconvolved estimators $\hat C_l$ can be expressed in terms of that of the convolved estimators 
$\tilde C_l$ as follows:

\begin{eqnarray}
\left (\begin{array}{cc}
        \langle \delta \hat C_L^{X} \delta \hat C_{L'}^{X}\rangle & \langle \delta \hat C_L^{X} \delta \hat C_{L'}^{Y}\rangle   \\
        \langle \delta \hat C_L^{Y} \delta \hat C_{L'}^{X}\rangle & \langle \delta \hat C_L^{Y} \delta \hat C_{L'}^{Y}\rangle   \\
       \end{array}
\right ) =
\left (\begin{array}{cc}
   \mathrm M^{XX}_{Ll} & \mathrm M^{XY}_{Ll} \\
   \mathrm M^{YX}_{Ll} & \mathrm M^{YY}_{Ll} 
  \end{array}
\right )^{-1}
\left (\begin{array}{cc}
        \langle \delta \tilde C_l^{X} \delta \tilde C_{l'}^{X}\rangle & \langle \delta \tilde C_l^{X} \delta \tilde C_{l'}^{X}\rangle   \\
        \langle \delta \tilde C_l^{Y} \delta \tilde C_{l'}^{X}\rangle & \langle \delta \tilde C_l^{Y} \delta \tilde C_{l'}^{Y}\rangle   \\
       \end{array}
\right )
{\left (\begin{array}{cc}
   \mathrm M^{XX}_{L'l'} & \mathrm M^{XY}_{L'l'} \\
   \mathrm M^{YX}_{L'l'} & \mathrm M^{YY}_{L'l'} 
  \end{array}
\right )^{-1}}^T
\end{eqnarray}

\noindent 
In deriving these results it is assumed that the coverage of the sky is near complete.
This will mean that the windows associated with the various couplings are sharper than any features in the power spectra.  The shape of the mask and the noise covariance properties are quite general at this stage.If the $C_l$s of individual
data sets are known from independent estimations then cross spectra deconvolution of the cross-spectra $C_l^{X,X}$  can be
simply written as:

\begin{equation}
\langle \hat C_L^{X,Y} \hat C_{L'}^{X,Y}  \rangle = \sum_{ll'} [{\mathrm M^{X,Y}}]_{Ll}^{-1} \langle \tilde C_l^{X,Y} \tilde C_{l'}^{X,Y} \rangle 
[ \mathrm M^{\rm\rY} ]_{L'l'}^{-1}.
\end{equation}

\begin{figure}
\begin{center}
{\epsfxsize=10. cm \epsfysize=5. cm {\epsfbox[23 439 590 715]{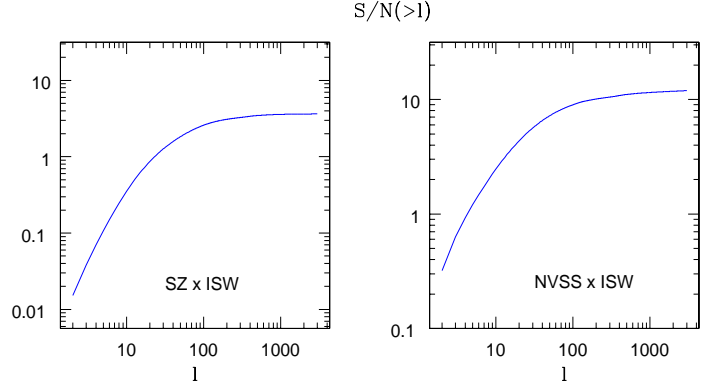}}}
\end{center}
\caption{The cumulative signal to noise for ISW cross LSS (right panel) and SZ cross ISW (left panel)
are plotted as a function of $l$. The results are obtained by using the covariances of $C_l$s 
plotted in previous plot. The $S/N$ will degrade linearly with $f_{sky}$. Plots correspond to
$f_{sky} =1$.} 
\label{fig:s2n_ps}
\end{figure}

\n
In the limiting situation when the survey area covers almost the entire sky these equation takes
a much simpler form which are in common use in the literature. If $f_{sky}$ is the fraction of the sky
covered then one can write:

\begin{equation}
\langle \delta C_l^{XY} \delta {\cal C}_{l'}^{XY} \rangle = 
 f_{sky}{1\over (2l+1)}[{\cal C}_l^X {\cal C}_l^Y + ({{\cal C}_l^{XY}})^2 ]\delta_{ll'}; \qtwo
\langle \delta {\cal C}_l^{XY} \delta {\cal C}_{l'}^{X,Y} \rangle = 
f_{sky}{2\over (2l+1)}({\cal C}_l^{X,Y} {\cal C}_l^{XY} )\delta_{ll'}.
\end{equation}

\n
The $C_l$s in these expression are the total ${\cal C}_l = {\cal C}_l^S + {\cal C}_l^N$ which takes contribution from
both signal and noise ${\cal C}_l$s.

\section{The Pseudo-${\cal C}_{\ell}$ Estimator for mixed Bispectrum Analysis}

The statistics of temperature fluctuations in the sky are very nearly Gaussian, but small departures from Gaussianity can put
constraints on early universe scenarios. Secondary non-Gaussianity 
on the other hand can provide valuable information to distinguish structure formation scenarios, and
when used with constraints from the power spectrum it can be a very valuable tool.
However estimation of the bispectrum for each triplet of harmonics modes can be difficult
to perform numerically.   \cite{MuHe09} introduced a limited data compression method for 3-point functions which reduces the data to a single function (the skew spectrum), and which can be made optimal for estimating a bispectrum form.  In the same spirit, we define a pseudo-skew spectrum for three arbitrary fields defined on a cut sky, and show how it is related to the skew spectrum on the uncut sky.    The PCL-based approach described here is not optimal, however, it can be made optimal
with suitable choice of weights.

Let us assume that we have three fields which are defined over the observed sky. The product
of two of these fields as $\rX(\hat \Omega)\rY(\hat \Omega)$ has an associated mask which, we denote as $w_A(\Omega)$, and which is a product of two masks associated with the
individual fields. Analogously, the third field $\rZ(\Omega)$ is observed with a mask $w_B(\Omega)$. 

From the harmonic transforms of these fields we study the skew spectrum
and express it in terms of the mixed bispectra of fields X, Y and Z, $B_{l_1l_2l_3}^{XYZ}$. 
We develop this generally, but the results we derive will be useful for the study of
primordial non-Gaussianity.  Here we consider a single field (the CMB), but it is a field with contributions from various components, and the skew spectrum contains terms from various triplet of different (or repeated) fields.

\subsection{Estimator for $\tilde {\cal C}_l ^{\rX\rY,\rZ}$}

\subsubsection{All-Sky Analysis}

\n
We start by introducing the power spectrum $\tilde {\cal C}_l ^{\rX\rY,\rZ}$ associated with
the cross-correlation of the product map $\rX(\oh)\rY(\oh)$ and $\rZ(\oh)$. In the absence of
sky-cuts and instrumental noise we can write:

\begin{equation}
\hat {\cal C}_l ^{\rX\rY,\rZ} = {1 \over 2l+1} \sum_m  a_{lm}^{XY} a_{lm}^{Z*}
\end{equation}
where $a_{lm}^{XY}$ is the spherical harmonic transform of $XY$.

Assuming homogeneity and isotropy, the correlation function $\mathcal C(\Omega, \Omega')$ of $\mathrm X(\Omega)Y(\Omega)$ and $Z(\Omega)$ can be written in terms of $\hat {\cal C}_l ^{\rX\rY,\rZ}$ 

\begin{equation}
\mathcal C(\Omega, \Omega') \equiv \langle \rX(\Omega)\rY(\Omega)\rZ(\Omega')\rangle = 
\sum_{l_1m_1, l_2m_2} \langle a_{l_1m_1}^{\rX\rY} a_{l_2m_2}^{\rZ}\rangle Y_{l_1m_1}(\Omega) Y_{l_2m_2}(\Omega') = { 1 \over 4 \pi }\sum_l (2l+1)P_l(\cos(\oh \cdot\oh')) \hat C_l ^{\rX\rY,\rZ}.
\end{equation}

\n
Here $P_l(\oh \cdot\oh')$ is a Legendre polynomial of order $l$.
The three-point correlation function in the harmonic domain can similarly be used to
introduce the mixed bispectrum $B_{l_1l_2l_3}^{\rX\rY\rZ}$ for the related fields.  Assuming statistical isotropy,

\begin{equation}
\langle a^{\rX}_{l_1m_1}a^{\rY}_{l_2m_2}a^{\rZ}_{l_3m_3}\rangle = 
\left ( \begin{array}{ c c c }
     l_1 & l_2 & l_3 \\
     m_1 & m_2 & m_3
  \end{array} \right) B_{l_1l_2l_3}^{\rX\rY\rZ}.
\end{equation}

\n
Our aim is to compute the cross-correlation power spectra of the product field 
$\mathrm X(\Omega)Y(\Omega)$ and $Z(\Omega)$. Using a harmonic decomposition
we can relate the multipoles $a^{\rX\rY}_{lm}$ with multipoles $a^{\rX}_{l'm'}$ and 
$a^{\rY}_{l''m''}$:

\begin{eqnarray}
a^{\rX\rY}_{lm} &=& \int d \hat {\Omega} Y_{lm}^*(\hat { \Omega})\rX(\hat {\Omega})\rY(\hat {\Omega}) 
=  \sum_{l'm'} \sum_{l''m''} a^{X}_{l'm'} a^{Y}_{l''m''} \int  d  \hat\Omega Y_{lm}^*( \hat\Omega)Y_{l'm'}( \hat\Omega)Y_{l''m''}( \hat\Omega)  \nonumber\\
&=& \sum_{l'm'} \sum_{l''m''} a^{X}_{l'm'} a^{Y}_{l''m''} \sqrt {(2l_1 +1)(2l_2+1)(2l_3+1) \over 4\pi  }
\left ( \begin{array}{ c c c }
     l & l' & l'' \\
     0 & 0 & 0
  \end{array} \right)
\left ( \begin{array}{ c c c }
     l & l' & l'' \\
     m & m' & m''
  \end{array} \right).
\end{eqnarray}

\n
Contracting with the multipole of the remaining field $a_{l'm'}^{\rZ}$ we can see that
it directly probes the mixed bispectrum associated with these three different fields 
\citep{Cooray1}.

\begin{eqnarray}
&& \langle a_{lm}^{\rX\rY}a_{l'm'}^{Z*} \rangle \equiv {\cal C}_l^{XY,Z} \delta_{ll'} \delta_{mm'};\qquad
  {\cal C}_l^{XY,Z} = \sum_{l_1,l_2} B_{ll_1l_2}^{\rX\rY\rZ} \sqrt{(2l_1+1)(2l_2+1)\over 4 \pi (2l +1) }
\left ( \begin{array}{ c c c }
     l_1 & l_2 & l _3 \\
     0 & 0 & 0
  \end{array} \right)
\end{eqnarray} 

\n 
Since the bispectrum is determined by triangular configuration in the multipole space $(l_1,l_2,l)$, the power spectrum $C_l$ defined above captures information about all possible triangular configuration when one of its sides is fixed at length $l$. However this data compression is not done optimally as it does not weight the contributions from each bispectrum 
components with their inverse variance. The error covariance matrix can be
computed exactly and depends on higher-order moments of signal and noise, as well as
their cross-correlations.

\subsubsection{Partial sky Coverage}

\n
It is possible to extend the above result to take into account partial sky coverage. 
Assuming the composite map $X(\hat \Omega)Y(\hat \Omega)$ is masked with 
arbitrary mask $w_A(\hat \Omega)$ and the map $Z(\hat \Omega)$ is masked with 
$w_B(\hat \Omega)$ we can write the cut-sky multipoles  $\tilde a^{\rX\rY}_{lm}$
and $\tilde a^{\rZ}_{lm}$ in terms of their all sky counterparts as well as
the multipoles associated with the mask multipoles follows:

\begin{eqnarray}
\tilde a^{\rX\rY}_{lm} = \int X(\hat \Omega)Y(\hat \Omega) w_A(\hat \Omega) Y^*_{lm}(\hat \Omega) d \hat\Omega; 
\qtwo a^{\rZ}_{lm} = \int Z(\hat \Omega) w_B(\hat\Omega) Y^*_{lm}(\hat \Omega) d \hat\Omega 
\end{eqnarray}

\begin{eqnarray}
&&\tilde a_{lm}^{XY} = \sum_{l_1m_1} \sum_{l_2m_2} \sum_{l_Am_A} a^{X}_{l_1m_1}a^{Y}_{l_2m_2}w_{l_Am_A} 
\int d \hat \Omega Y_{l_1m_1}({\hat \Omega})Y_{l_2m_2}({\hat \Omega}) Y_{l_Am_A}({\hat \Omega}) \nonumber \\
&&\tilde a_{lm}^{Z} =  \sum_{l_3m_3} \sum_{l_Bm_B} a^{Z}_{l_3m_3}w_{l_Bm_B}
\int d \hat \Omega Y_{l_3m_3} ({\hat \Omega}) Y_{l_Bm_B}({\hat \Omega}) .
\end{eqnarray}

\noindent
Here $B_{l_1l_2l_3}$ is the angle-averaged bispectrum and the functions $b_{l_1}$
represent the effects of pixellisation as well as beam smoothing. For partial sky coverage one can 
obtain after tedious but straight forward algebra:

\beqa
\tilde {\cal C}_l^{\rX\rY,\rZ} = \sum_{l'}{2l'+1 \over 4\pi} \sum_{l''}{(2l''+1) \over 4\pi}
\left ( \begin{array}{ c c c }
     l & l' & l'' \\
     0 & 0 & 0
  \end{array} \right)^2
|w_{l''}|^2  \sum_{l_1,l_2} B_{l'l_1l_2}^{XYZ} \sqrt{(2l_1+1)(2l_2+1) \over (2l'+1) 4 \pi}
\left ( \begin{array}{ c c c }
     l_1 & l_2 & l' \\
     0 & 0 & 0
  \end{array} \right)
   \equiv M_{ll'} \hat {\cal C}_{l'}^{XY,Z}.
\eeqa

\noindent
This is one of the important results in this paper.  It shows how the pseudo-skew spectrum is related to the all-sky skew spectrum, and is a computationally-efficient way to estimate the latter.  By suitable choice of weight functions it can be made optimal.   It is valid for a completely general mask where $w_{l}$ represents the spherical
transform of the mask.

The transformation matrix  $M_{ll'}$ used here is the same as that we 
introduced for the recovery of cross power spectra. The power spectra 
associated with the mask $w_l$, plays the same role in construction 
of $M_{ll'}$ For simplicity we have assumed that different data sets have 
the same mask but it is trivial to generalise for two different masks.
A detailed comparison of level of sub-optimality will be
compared with numerical simulations in an accompanying paper. This analysis
is complementary to work by \citet{ChSz06} where a similar sub-optimal estimator
was used to study non-Gaussianity.

\section{CMB Secondary Bispectrum}

The formalism developed so far is quite general and can handle mixed bispectra
of different kinds. The main goal was to relate the skew spectrum with the corresponding mixed bispectrum. To make
concrete predictions we need to consider a specific form for the bispectrum.
Following  \citet{SperGold99}, \citet{GoldbergSpergel99} and \citet{CoorayHu} we expand the observed temperature 
anisotropy $\delta T(\Omega)$ in terms of the primary anisotropy $\delta T_{\mathrm {P}}$,
and due to lensing of primary, $\delta T_{\mathrm {L}}$, and the other
secondaries from coupling large-scale structure, $\delta T_{\mathrm {S}}$.

\begin{equation}
\delta T(\oh) = \delta T_{\mathrm {P}}(\oh) + \delta T_{\mathrm {L}}(\oh) + \delta T_{\mathrm {S}}(\oh) .
\end{equation}

\n
Expanding the respective terms in spherical harmonics we can write:

\begin{equation}
\delta T_{\mathrm {P}}(\oh) \equiv \sum_{lm} a_{lm} Y_{lm} (\oh); ~~~~
\delta T_{\mathrm {L}}(\oh) \equiv \sum_{lm} a_{lm} \nabla \Theta(\oh) \cdot \nabla T_{\mathrm {S}}(\oh); ~~~~
\delta T_{\mathrm {S}}(\oh) \equiv \sum_{lm} b_{lm} Y_{lm} (\oh).
\end{equation}

\n
The harmonic coefficients $b_{lm}$ are associated with the expansion of the
secondary anisotropies $\delta T_{\mathrm {S}}(\oh)$. The secondary 
bispectrum for the CMB then takes contributions from many products of $P,L,S$ terms.  For example, one term arises from products of 
$\delta T_{\mathrm {P}}\delta T_{\mathrm {L}}\delta T_{\mathrm {S}}$:

\begin{eqnarray}
B_{l_1l_2l_3}^{PLS} && \equiv \sum_{m_1m_2m_3} \left ( \begin{array}{ c c c }
     l_1 & l_2 & l_3 \\
     m_1 & m_2 & m_3
  \end{array} \right) \int \left \langle \delta T_P(\oh_1) \delta T_L(\oh_2) \delta T_S(\oh_3) \right \rangle
Y_{l_1m_1}(\oh_1) Y_{l_2m_2}(\oh_2) Y_{l_3m_3}(\oh_3) d \oh_1 d \oh_2 d \oh_3 \nonumber \\
&& \equiv \sum_{m_1m_2m_3} \left ( \begin{array}{ c c c }
     l_1 & l_2 & l_3 \\
     m_1 & m_2 & m_3
  \end{array} \right) \langle (\delta T_{\mathrm {P}})_{l_1m_1} (\delta T_{\mathrm {L}})_{l_2m_2}
(\delta T_{\mathrm {S}})_{l_3m_3}  \rangle  .
\end{eqnarray}

\n
It is possible to invert the relation using isotropy of the background cosmology:

\begin{equation}
\langle (\delta T_{\mathrm {P}})_{l_1m_1} (\delta T_{\mathrm {P}})_{l_2m_2}
(\delta T_{\mathrm {P}})_{l_3m_3}  \rangle = \left ( \begin{array}{ c c c }
     l_1 & l_2 & l_3 \\
     m_1 & m_2 & m_3
  \end{array} \right) B_{l_1l_2l_3}.
\end{equation}

\n
Explicit calculations, detailed in \citet{GoldbergSpergel99} and \citet{CoorayHu}, found the mixed bispectrum to be of the 
following form:

\begin{equation}
B_{l_1l_2l_3}^{PLS} = -\left \{  b_{l_3} c_{l_1} { l_2(l_2+1) - l_1(l_1+1) - l_3(l_3+1) \over 2 }+cyc.perm. \right \} \sqrt {(2l_1 +1)(2l_2+1)(2l_3+1) \over 4\pi  }\left ( \begin{array}{ c c c }
     l_1 & l_2 & l_3 \\
     0 & 0 & 0
  \end{array} \right) \equiv b_{l_1l_2l_3}I_{l_1l_2l_3},
\label{eq:bispec_intro}
\end{equation}

\n
where we have defined the reduced bispectra $b_{l_1l_2l_3}$, which is useful 
in certain context \citep{Bartolo06}, and the additional geometrical
factor, which originates from the integral involving three spherical harmonics
over the entire sky:

\begin{equation}
I_{l_1l_2l_3} \equiv \sqrt {(2l_1 +1)(2l_2+1)(2l_3+1) \over 4\pi  }\left ( \begin{array}{ c c c }
     l_1 & l_2 & l_3 \\
     0 & 0 & 0
  \end{array} \right) .
\end{equation}

\n
The cross-correlation power-spectra appearing in the above expression denotes the coupling of lensing
with a specific form of secondary non-anisotropy (see e.g. \citet{CoorayHu}).

\begin{equation}
\langle \Theta_{l'm'} b_{lm} \rangle = b_l \delta_{ll'}\delta_{mm'}.
\end{equation}
             
\n
The bispectrum contains all the information at the three-point correlation function level
and can be reduced to one-point skewness or the two-point collapsed correlation function
or the associated power-spectra, the skew spectrum.
Here we have considered the secondaries of the CMB, however the analysis holds if external 
tracers such as the radio galaxy surveys such as NVSS or 21cm observations are used instead.
In the next few subsections we will discuss the problem of estimation of the skew spectrum in a nearly optimal way. 
This will lead to a discussion of development
of optimal techniques in subsequent sections. We will also tackle the problem of 
joint estimation of several bispectra and associated estimation errors. As is known
and will be discussed in the following sections that the estimation of CMB bispectra is
similar and related to the case of lensing reconstruction of the CMB sky \citep{SmZaDo00}.

\section{Estimation of skew spectra }

The problem of estimation of the skew spectrum is
very similar to that of the primary CMB bispectrum. There has been a recent surge in activity in this area, driven by 
the claim of detection of non-gaussianity in the WMAP data release (see e.g.\citet{YaWa08,Yadav08,YKW}). Different techniques 
were developed which
introduce various weighting schemes in the harmonic domain to make the method optimal (i.e. saturates the Cramer-Rao bound).
Maps are constructed by weighting the observed CMB sky with $l$-dependent weights
obtained from inflationary theoretical models. These weighted maps are then used to compute 
one-point quantities which are generalisation of skewness and can be termed mixed skewness. 
These mixed skewness measures are useful estimators of $f_{NL}$ parameters. A more general treatment
was provided in \cite{SmZa06,SmZaDo00,SmSeZa09} who took into account mode-mode coupling
in an exact way with the use of proper inverse-covariance weighting of harmonic modes.

Recent work by \cite{MuHe09} has improved the situation by focussing directly on
the skew spectrum. Their technique does not compress
all the available information in the bispectrum into a single number but provides a power-spectrum
which depends on the harmonic wavenumber $l$. This method has the advantage of being able to separate
various contributions as they will have different dependence on $l$s, thus allowing an assessment of whether any non-gaussianity is primordial or not. 
In this section we compute the contaminating secondary bispectrum contributions from lensing-secondary coupling.

\subsection{Bispectra without line of sight integration involving ISW-lensing RS-lensing and SZ-lensing}

The study of the bispectrum related to secondary anisotropy (see \cite{sethcoo}
for more details and analytical modelling based on halo model which we use here)
is arguably as important as that generated by the primary anisotropy. Primary non-Gaussianity
in simpler inflationary models is vanishingly small
\citep{Salopek90,Salopek91,Falk93,Gangui94,Acq03,Mal03}; see \citet{Bartolo06}
and references therein for more details.
However, variants of simple inflationary models such as multiple
scalar fields \citep{lindemukha,Lyth03}, features in the inflationary potential, 
non-adiabatic fluctuations, non-standard
kinetic terms, warm inflation \citep{GuBeHea02,Moss}, or
deviations from Bunch-Davies vacuum can all lead to a much higher level of non-Gaussianity.  

Early observational work on the bispectrum
from COBE \citep{Komatsu02} and MAXIMA \citep{Santos} was followed by much more 
accurate analysis with WMAP \citep{Komatsu03,Crem07a,Spergel07}.
 The primary bispectrum encodes information about inflationary dynamics
and hence can constrain various inflationary scenarios, where as
the secondary bispectrum will provide valuable information regarding
the low-redshift universe and constrain structure formation scenarios. 
These bispectra are generated because of the cross-correlation effect 
of lensing due to various intervening materials and the
secondary anisotropy such as the Sunyaev-Zeldovich effect due to
inverse Compton scattering of CMB photons from hot gas 
in intervening clusters.

\begin{equation}
b_{l_1l_2l_3} = -{1 \over 2} \big [ \left \{ (l_2(l_2+1)-l_1(l_1+1)-l_3(l_3+1) \right \} \myC_{l_1}^Sb_{l_3} + 
{\rm cyc.perm.}~ \big ].
\label{eq:bi_komatsu}
\end{equation}

\n
The power spectrum $\myC^S_l$ is the unlensed power spectrum of the CMB anisotropy.
We have introduced the subscript $S$ to distinguish it from the $C_l$s that appear in the denominator
which take contribution from the instrumental noise from signal to noise computation point of view ($C_l=C_l+N/b(l)^2$, where $N$ is the
instrumental noise and $b(l)$ is the beam function in multipole space). 
We define the different fields which are constructed from underlying harmonics
and corresponding ${\cal C}_l$s. These will be useful for constructing an unbiased near optimal
estimator.

\begin{eqnarray}
A^{(1)}_{lm} = {a_{lm}\over {\cal C}_{l}}{\cal C}^S_{l}; \qtwo B^{(1)}_{lm} = l(l+1) {a_{lm}\over{\cal C}_{l}}; \qtwo { C}^{(1)}_{lm} = {a_{lm} \over {\cal C}_{l}}b_l \nonumber \\
A^{(2)}_{lm} = -l(l+1){a_{lm}\over {\cal C}_{l}}{\cal C}^S_{l}; \qtwo B^{(2)}_{lm} =  {a_{lm}\over{\cal C}_{l}}; \qtwo { C}^{(2)}_{lm} = {a_{lm} \over {\cal C}_{l}}b_l \nonumber \\
A^{(3)}_{lm} = {a_{lm}\over {\cal C}_{l}}{\cal C}^S_{l}; \qtwo B^{(3)}_{lm} =  {a_{lm}\over{\cal C}_{l}}; \qtwo {C}^{(3)}_{lm} = l(l+1){a_{lm} \over {\cal C}_{l}}b_l 
\end{eqnarray}

\n
The corresponding fields that we construct are
 $A^{(i)}(\hat \Omega) \equiv \sum_{lm}Y_{lm}(\hat\Omega)A^{(i)}_{lm}$, and in an analogous manner 
$B^{(i)}$ and  $C^{(i)}$. The optimised skew spectrum in the presence of all-sky coverage and homogeneous 
noise can now be written as:

\begin{equation}
{\cal C}_l^{2,1}  =  
{1 \over 2l + 1} \sum_{m} \sum_i {\rm Real} \{ (A^{(i)}(\oh)B^{(i)}(\oh))_{lm} C^{(i)*}(\oh)_{lm} \} + cyc.perm.
\end{equation}

\n
The cyclic terms that are considered here will have to constructed likewise from the corresponding
terms in the expression for the reduced bispectrum discussed above \ref{eq:bi_komatsu}. 
The linear-order correction terms which needs to be included in the absence of spherical symmetry
due to presence of cuts to avoid the galactic foreground and the inhomogeneous noise can be
written as:

\begin{equation}
\hat {\cal C}_l^{2,1} = { 1 \over f_{sky}} \sum_i \left [ \tilde {\cal C}_l^{AB,C} - {\cal C}_l^{A \langle B,C \rangle} - {\cal C}_l^{B \langle A, C \rangle} - {\cal C}_l^{\langle AB \rangle,C} \right ]^{(i)} + cyc.perm.
\end{equation}

\n
The terms without averaging such as $\tilde C_l^{AB,C}$ are direct estimates from the 
observed partial sky with inhomogeneous noise. The Monte Carlo corrections such as
$ C_l^{A \langle B,C \rangle}$  are constructed by cross-correlating the product of the
observed map $A$ and a Monte Carlo map $B$ with a Monte Carlo map $C$ and then taking
an ensemble average over many realisations. The denominator $f_{sky}$, which represent the
fraction of the sky covered, is introduced to correct for the effect of partial 
sky coverage. This is an approximate way to treat the mode-mode coupling due to partial
sky coverage and known to be a good approximation for higher $l$.
The skewness associated with this form of 
bispectra can be expressed as a weighted sum of the corresponding $C_l$s:

\begin{equation}
\hat S = \sum_l (2l+1) \hat {\cal C}_l^{2,1} = \sum_{ll_1l_2}{\hat B_{ll_1l_2}B_{ll_1l_2} \over {C_lC_{l_1}C_{l_2}}}.
\end{equation}

\n
Constructing such weighted  maps clearly can be seen as a way to construct a matched-filter
estimator for the detection of non-Gaussianity. It is optimally weighted by the inverse cosmic
variance and achieves maximum response when the observed non-Gaussianity matches 
with a specific theoretical input. The skew spectrum also allows for analysis of more than one specific
type of non-gaussianity from the same data - allowing a joint analysis to determine cross-contamination
from various contributions.

\subsection{Bispectra involving line of sight integration: The Ostriker-Vishniac 
effect and its correlation with other secondary anisotropies }

\n
Another set of secondary bispectra involving any of the Ostriker-Vishniac (e.g. see \cite{JaffeKamion})
effect, SZ thermal effect, or the kinetic SZ effect \citep{Cooray2} or a combination of these  have 
the following form of reduced bispectrum \citep{CoorayHu}, which involves a line of sight integration along $r$:

\begin{equation}
b_{l_1l_2l_3} = \int dr~f_{l_1}(r)g_{l_2}(r) + {\rm cyc.perm.}  
\end{equation}

\n
The construction of weighted maps follow the same principle with the use of kernels $f_l(r)$ and $g_l(r)$ that
are associated with  any of the scattering secondaries that involve a line of sight integration.
Numerical implementation of line of sight will naturally have to deal with an optimal method to include the 
quadrature. Defining

\begin{equation}
A_{lm}(r) = {a_{lm} \over {\cal C}_{l}} f_{l}(r);~~~~  B_{lm}(r) = {a_{lm} \over {\cal C}_{l}} g_{l}(r);
~~~ {C}_{lm}(r) = { a_{lm} \over {\cal C}_{l}}.
\end{equation}

\n
$A$ and $B$ are fields constructed from the generic function represented by $f_l$ and $g_l$.  Following \cite{MuHe09} we construct
\begin{equation}
\hat {\cal C}_l^{2,1}(r) = {1 \over 2l+1} \int dr ~r \sum_{m} {\rm Real} \{ (AB)_{lm}(\oh,r) C(\oh,r)_{lm} \} +cyc.perm.
\end{equation}
and from this compute the skew-spectrum:
\begin{equation}
\hat{\cal C}_l^{2,1} = \int dr ~\hat {\cal C}_l^{2,1}(r).
\end{equation}

\n
This is the generalisation of the all-sky estimator of the skew spectrum of \cite{MuHe09}, but for three distinct fields.

The corresponding one-point skewness can be written as

\begin{equation}
\hat S = \sum_l(2l+1){\hat \myC}_l^{2,1} = \sum_{ll_1l_2} {B_{ll_1l_2}\hat B_{ll_1l_2}\over {\cal C}_{l}{\cal C}_{l_2}{\cal C}_{l_3}}.
\end{equation}

\n
In the next section we consider the contamination of the primary skew spectrum by 
secondary non-Gaussianity from point sources. As before we have absorbed the beam 
in the harmonic coeeficients of the data vector $a_{lm}$. As before, ${\cal C}_l$s also
take contribution from the noise $\myC^N_l$ as well as from the theoretical CMB powerspectra 
$\myC^S_l$, i.e.  ${\myC}_l = \myC_l^S + \myC^N_l$

\subsection{General Expression}

From the examples above, its clear that from a very general consideration if the reduced bispectrum can be decomposed in such a way
it consists of terms, which can be used to construct fields such as $A^{(i)}, B^{(i)}$ and $C^{(i)}$ (not necessarily of
a specific form) a skew-spectrum can always be constructed by similar manipulation. In certain cases the $C_l^{2,1}(r)$
might actually also have radial dependence, in which case a line of sight integration needs to be performed to match
observations.

\begin{equation}
\hat C_l^{2,1} = {1 \over (2L+1)} \sum_{l'l''} \sum_{ij} {(2l' +1)(2l''+1) \over 144 \pi} \left ( \begin{array}{ c c c }
     l' & l'' & L \\
     0 & 0 & 0
  \end{array} \right)^2 { 1 \over {\cal C}_{L}} {1 \over {\cal C}_{l'}}{1 \over {\cal C}_{l''}} 
\left [ A^{i}_{l_1} B^{i}_{l_2}C^{i}_{l_3} + \rm {cyc.perm.} \right ]\left [ A^{j}_{l_1} B^{j}_{l_2}C^{j}_{l_3} + \rm {cyc.perm.} \right ]
\end{equation}

\subsection{Cross-contamination from Point Sources and Primary non-Gaussianity}

\n
The bispectra associated with point sources is modelled as $b^{ps}_{l_1l_2l_3} = const.$
The constant depends on the flux limit. More complicated modelling which incorporates certain
aspects of halo models can be used for better accuracy \cite{SerCoo08}.

\begin{equation}
S  = \sum_{l_1l_2l_3} {B_{l_1l_2l_3}^{ps}B_{l_1l_2l_3}^{sec}\over {\cal C}_{l_1}{\cal C}_{l_2}{\cal C}_{l_3}}.
\end{equation}

\n
Similarly given a model of primary non-Gaussianity one can construct a theoretical
model for computation of $B_{ll_1l_2}^{prim}$ (see Munshi \& Heavens 2009 for more
about various models and construction of optimal estimators). While study of primary
non-Gaussianity is important in its own right for the study of secondaries they
can confuse the study.

\begin{equation}
S  = \sum_{l_1l_2l_3} {B_{l_1l_2l_3}^{prim}B_{l_1l_2l_3}^{sec}\over {\cal C}_{l_1}{\cal C}_{l_2}{\cal C}_{l_3}}
\end{equation}

\n
Similar results hold at the level of the skew spectrum.
A more general treatment based on Fisher analysis of multiple bispectra is
presented in the subsequent sections.

In addition to various sources mentioned above, second-order corrections to the gravitational potential through gravitational instability too can also act as a source of secondary non-Gaussianity \citep{MuSoSt95}.

\section{Optimised Analysis of Mixed Bispectra}

Starting from \citet{Babich} a complete analysis of bispectrum in the presence of partial sky coverage
and inhomogeneous noise was developed by various authors \citep{Babich,Crem06,Yadav08}.
A specific form for a bispectrum estimator was introduced which is both unbiased and 
optimal. This was further developed and used by \citet{SmZaDo00} for
lensing reconstruction and by \citet{SmZa06} for general bispectrum analysis.
for one-point estimator for $f_{NL}$. The analysis depends on
finding suitable inverse cosmic variance weighting $C^{-1}$ of
modes. It deals with mode-mode coupling in an exact way. In a recent
work \citet{MuHe09} further extended this analysis by incorporating
two-point statistics or the skew spectrum
which we have already introduced above. We generalise their results in this
work for the case of mixed bispectra for the case of both one-point and 
two-point studies involving three-way correlations. The analytical results 
presented here are being kept as general as possible.
However in the next sections we specialise them to individual cases
of lensing reconstruction and the mixed bispectrum associated with lensing
and the SZ effect as concrete examples.

\subsection{One-point Estimator: Mixed Skewness $\langle X(\oh)Y(\oh)Z(\oh)\rangle$}

\n
We are interested in constructing an optimal and unbiased estimator for
the estimation of mixed skewness $\langle X(\oh)Y(\oh)Z(\oh) \rangle$.
The fields $X(\Omega) =\sum_{lm}X_{lm}Y_{lm}(\oh)$ and similarly for $Y$ and $Z$, are defined over the entire sky, though
observed with a mask and nonuniform noise coverage. The non-uniform coverage
imprints a mode-mode coupling $\left [ C_{XX} \right ]^{-1}_{l_1m_1,l_2m_2}$ in the observed
multipoles of a given field in the harmonic space $\langle X_{l_1m_1}X_{l_2m_2}\rangle$. 
For the construction of the optimal estimator it will be useful to define $\tilde X_{l_1m_1}$
as

\begin{equation}
\tilde X_{l_1m_1} \equiv \left [ C_{XX} \right ]^{-1}_{l_1m_1,l_am_a} X_{l_am_a}.
\end{equation}

\n
Here $\tilde X_{lm}$ represents the harmonics of the data $X$ with inverse 
covariance weighting. 
Next we need to deal with the covariance matrix of the modes $\tilde X_{l_1m_1}$ 
in terms of that of $X_{l_1m_1}$
The auto covariance matrix for $X$, $C_{XX}$, and that of $\tilde X$, $\tilde C_{XX}$ are
related by the following expression: 

\begin{equation}
[\tilde C_{XX}]_{l_am_a,l_bm_b} \equiv \langle \tilde X_{l_am_a} \tilde X_{l_bm_b} \rangle = [C^{-1}_{XX}]_{l_am_a,l_bm_b}  .
\end{equation}

\n
Similarly, the cross-covariance for two different fields $\tilde X$ and $\tilde Y$ with inverse
variance weighting, in harmonic space can be written as:

\begin{equation}
 \tilde C_{XY} \equiv \langle \tilde X_{l_am_a} \tilde Y_{l_bm_b} \rangle = C^{-1}_{XX} C_{XY}  C^{-1}_{YY}.
\end{equation}

\n
The estimator that we construct will be based on functions $\hat Q[\tilde X, \tilde Y, \tilde Z]$
which depends on the input fields, and its derivatives w.r.t. the fields e.g. ${\partial \hat Q[\tilde X, \tilde Y] / \partial \tilde Z_{lm}}$. The derivatives are themselves a map with harmonics
described by the free indices $lm$, and are constructed out of two other maps. The function $\hat Q$
on the other hand is an ordinary number which depends on all three input functions and lacks
free indices.

\begin{eqnarray}
&&\hat Q[\tilde X, \tilde Y, \tilde Z] \equiv {1 \over 6}\sum_{lml'm'l''m''} B_{ll'l''}^{XYZ}\left ( \begin{array}{ c c c }
     l & l' & l'' \\
     m & m' & m''
  \end{array} \right) \tilde X_{lm} \tilde Y_{l'm'} \tilde Z_{l''m''} \\
&& {\partial \hat Q[\tilde X, \tilde Y] \over \partial \tilde Z_{lm}} \equiv \sum_{lml'm'l''m''} B_{ll'l''}^{XYZ}\left ( \begin{array}{ c c c }
     l & l' & l'' \\
     m & m' & m''
  \end{array} \right) \tilde X_{l'm'} \tilde Y_{l''m''}.
\end{eqnarray}

\n
Similar expressions hold for other fields such as $\tilde X_{lm}$, $\tilde Y_{lm}$
Introducing a more compact notation $x_i$, where $x_1 = X, x_2 = Y, x_3 = Z$  we can write
the one-point estimator for the mixed skewness as:

\begin{equation}
\hat E[\tilde x_i ] = {1 \over F}\left \{ Q[\tilde x_i] - \sum_{i}{[\tilde {x}]^i_{l_am_a}} 
\langle \partial^i_{l_am_a}Q[\tilde x_i]\rangle \right \}.
\end{equation}

\n
This is a main result of the paper, generalising work by \citet{SmZa06} to mixed fields.

The ensemble averaging $\langle \rangle$ in the linear terms represents Monte-Carlo averaging using simulated 
non-Gaussian maps. The associated Fisher matrix (a scalar in this case) can be written in terms of the functions $Q[\tilde x_i]$,
its derivative and the cross-covariance matrices involving different fields.

\begin{equation}
F = \langle \partial^i_{l_am_a}Q[\tilde x_i] [C^{-1}]_{l_am_a,l_bm_b}^{ij} \partial^j_{l_bm_b}Q[\tilde x_j] \rangle
- \langle \partial^i_{l_am_a}Q[\tilde x_i] \rangle [C^{-1}]_{l_am_a,l_bm_b}^{ij} \langle \partial^j_{l_bm_b}Q[\tilde x_j] \rangle.
\end{equation}

\n
Here we have used the shorthand notation for $\langle x^i_{l_am_a} x^j_{l_bm_b} \rangle = [C^{-1}]_{l_am_a,l_bm_b}^{ij}$. In case of joint estimation of different bispectra from the same 
data sets we can extend the above discussion and write:

\begin{equation}
E[\tilde x_i ] = {F^{-1}}_{\alpha\beta}\left \{ Q^{\beta}[\tilde x_i] - {[\tilde {x}]^i_{l_am_a}} \langle \partial^i_{l_am_a}Q^{\beta}[\tilde x_i]\rangle \right \}.
\end{equation}

\n
Here the Fisher matrix $F_{\alpha\beta}$ encodes the inverse estimator covariance for
different mixed bispectra $B^{\alpha}$ and $B^{\beta}$, $\alpha$ and $\beta$ 
represents different types of bispectra recovered using the same data sets.

\begin{eqnarray}
F_{\alpha\beta} = && \sum B^{\alpha}_{l_1l_2l_3} B^{\beta}_{l_4l_5l_6}
\Big [ [\tilde C^{XX}]_{l_1m_1,l_4m_4}[\tilde C^{YY}]_{l_2m_2,l_5m_5}[\tilde C^{ZZ}]_{l_3m_3,l_6m_6}+ 
\nonumber\\
&& + [\tilde C^{XX}]_{l_1m_1,l_4m_4}[\tilde C^{YZ}]_{l_2m_2,l_6m_6}[\tilde C^{ZY}]_{l_3m_3,l_5m_5} +\nonumber \\
&& + [\tilde C^{XY}]_{l_1m_1,l_5m_5}[\tilde C^{YZ}]_{l_2m_2,l_6m_6}[\tilde C^{ZX}]_{l_3m_3,l_4m_4} + 
\tilde[\tilde C^{XZ}]_{l_1m_1,l_6m_6}[\tilde C^{ZY}]_{l_3m_3,l_5m_5}[\tilde C^{YX}]_{l_2m_2,l_4m_4} \Big ].
\end{eqnarray}

\n
The cyclic permutations here represent two additional terms with permutations of super-scripts 
$X,Y,Z$ along with associated subscripts. The Fisher matrix (which is a number in this particular case) 
for the mixed bispectrum in case of all-sky coverage and constant variance noise can be expressed as:

\begin{figure}
\begin{center}
{\epsfxsize=12. cm \epsfysize=6. cm {\epsfbox[28 521 590 715]{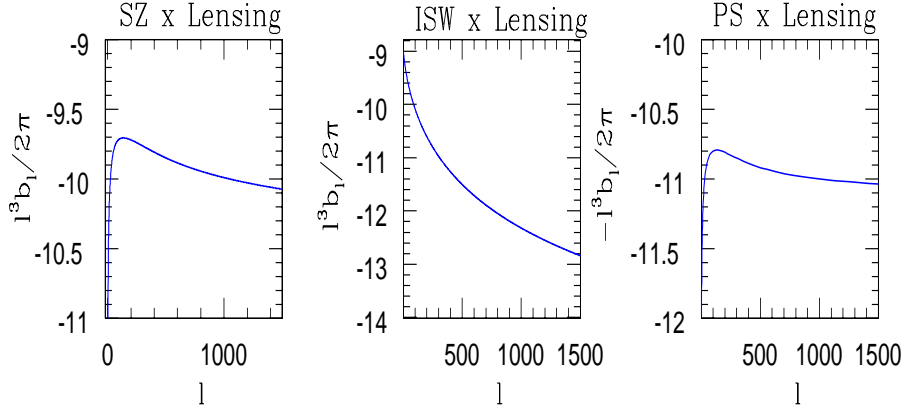}}}
\end{center}
\caption{The cross-spectra $b_l$ introduced in Eq.(\ref{eq:bispec_intro}) 
required for the construction of the bispectra is 
plotted for ISW cross lensing (right panel) and SZ cross lensing (left panel) bispectrum as a function of $l$. See text for details.}
\label{fig:bicls3}
\end{figure}

\begin{eqnarray}
F = {1 \over 6} \sum_{l_1l_2l_3} B_{l_1l_2l_3}^{XYZ} B_{l_1l_2l_3}^{XYZ}  \Big [
{1\over \myC_{l_1}^{XX}}  {1 \over \myC_{l_2}^{YY}\myC_{l_3}^{ZZ}}
+ {1 \over \myC^{XX}_l} \left ( {\myC_{l_1}^{YZ} \over \myC_{l_2}^{XX} \myC_{l_3}^{YY} } \right )^2 
+ 2{\myC_{l_1}^{YZ} \over \myC_{l_2}^{ZZ} \myC_{l_3}^{YY}}{ \myC_{l_1}^{ZX} \over \myC_{l_2}^{XX} \myC_{l_3}^{XX}}  
{\myC_{l_1}^{XY} \over \myC_{l_2}^{XX} \myC_{l_3}^{YY}}
\Big ]
\end{eqnarray}

\subsubsection{Special Case (A): Z=Y,~~  $\langle X(\oh)Y^2(\oh)\rangle$ }

\n
In certain practical situations we will encounter cases where two of the three fields are identical.
The corresponding Fisher matrix can be recovered by simply setting $Z=Y$. 

\begin{equation}
F = {1 \over 6 }\sum_{l_1l_2l_3} \left [ 2B_{l_1l_2l_3}^{XYY} B_{l_1l_2l_3}^{XYY} { \left ( 1 \over\myC_{l_1}^{XX} \right )}
{\left ( 1 \over \myC_{l_2}^{YY} \right )}^2 + 2B_{l_1l_2l_3}^{XYY} B_{l_3l_2l_1}^{XYY} {\left ( \myC_{l_1}^{XY}\over \myC_{l_1}^{XX}C_{l_1}^{YY} \right )^2}
{\left ( 1 \over \myC_{l_2}^{YY} \right )} \right].
\label{eq:skew_xyy}
\end{equation}

\subsubsection{Special Case (B): Z=Y=X, ~~ $\langle X^3(\oh) \rangle$}

\n
Finally, if we identify all three fields to recover the case ordinary or pure bispectrum 
corresponding to the case $X=Y=Z$.

\begin{equation}
F_{\alpha\beta} = {1 \over 6}\sum B^{\alpha}_{l_1l_2l_3} B^{\beta}_{l_4l_5l_6}
[C^{-1}]_{l_1m_1,l_4m_4}[C^{-1}]_{l_2m_2,l_5m_5}[C^{-1}]_{l_3m_3,l_6m_6}.
\end{equation}

\n
In the limit of all-sky coverage and constant variance noise the estimator reduces to:

\begin{equation}
F_{\alpha\beta} = {1 \over 6}\sum_{l_1l_2l_3} {B^{XYZ}_{l_1l_2l_3} B^{XYZ}_{l_1l_2l_3} 
\over \myC_{l_1}\myC_{l_2}\myC_{l_3} } 
\end{equation}

\n
For high $l$s a scaling $f_{sky}^{-1/2}$ is sufficient to describe the effect 
of partial sky coverage on the error covariance matrix.

\subsection{Two-point Estimators:  Mixed skew spectrum}
We begin by constructing the functions $Q_L$ and its derivative w.r.t. various input fields.
We use these to construct an optimal and unbiased estimator to correlate the field $X(\hat \Omega)$
with the product of two such fields $Y(\hat \Omega)Z(\hat \Omega)$. We consider the most
general possible case of the skew spectrum associated with the mixed bispectrum $B^{XYZ}$.
\begin{eqnarray}
&&\hat Q_L[\tilde X,\tilde Y, \tilde Z] \equiv \sum_{M} \tilde X_{LM} \sum_{l'm',l''m''} B_{Ll'l''}\left ( \begin{array}{ c c c }
     L & l' & l'' \\
     M & m' & m''
  \end{array} \right) \tilde Y_{l'm'} \tilde Z_{l''m''} \\
&& \partial_{lm}^X \hat Q_L[\tilde Y,\tilde Z] \equiv \delta_{Ll} \sum_{l'm',l''m''} B_{Ll'l''}\left ( \begin{array}{ c c c }
     L & l' & l'' \\
     m & m' & m''
  \end{array} \right) \tilde Y_{l'm'}\tilde Z_{l''m''} \\
&&  \partial_{lm}^Y \hat Q_L[\tilde X,\tilde Z] \equiv  \sum_{M} \tilde X_{LM} 
\sum_{l'm'} B_{Lll'}\left ( \begin{array}{ c c c }
     L & l & l' \\
     M & m & m'
  \end{array} \right) \tilde Z_{l'm'}; ~~
 \partial_{lm}^Z \hat Q_L[\tilde X,\tilde Y] \equiv  \sum_{M} \tilde X_{LM} 
\sum_{l'm'} B_{Lll'}\left ( \begin{array}{ c c c }
     L & l & l' \\
     M & m & m'
  \end{array} \right) \tilde Y_{l'm'}. 
\label{eq:xyy_qdq}
\end{eqnarray}
\n
While $Q_L$ is a number (cubic function of input maps) for a given $L$, the derivatives are maps which are 
quadratic in the input maps. The derivatives will be important in constructing the linear terms
which are important in reducing the variance of the estimator in an absence of spherical
symmetry, which is the case in the presence of inhomogeneous noise or partial sky coverage.
Using these expressions we can write down the optimised bispectra 
\begin{equation}
\hat E_L^{X,YZ}[x_i] = [N^{-1}]_{LL'}^{X,YZ} \left \{ Q_{L'}[\tilde x_i] - \sum_{i=1,2,3}[\tilde x_i]_{lm} \langle \partial^i_{lm} Q_{L'}[\tilde x_i]\rangle_{MC}) \right \}.\label{MainEstimator}
\end{equation}

\n
The normalisation matrix $N_{LL'}$  is related to the Fisher matrix $F_{LL'}= N^{-1}_{LL'}$,
and can be expressed as:

\begin{equation}
N_{LL'}^{X,YZ} = {1 \over 3} \left \langle \left \{ \partial_{l_1m_1}Q_L^i[\tilde x] \right \} [C^{-1}_{l_1m_1,l_2m_2}]^{ij}
\left \{ \partial_{l_2m_2}^jQ_{L'}[\tilde x]\right \} \right \rangle -{1 \over 3} \left \{ \left \langle \partial_{l_1m_1}^i Q_L[\tilde x] \right \rangle \right \} 
[C^{-1}_{l_1m_1,l_2m_2}]^{ij} \left \{ \left \langle \partial^j_{l_2m_2}Q_{L'}[\tilde x] \right \rangle \right \}.
\end{equation}

\n
Finally the Fisher matrix can be written as:

\begin{eqnarray}
F_{LL'}  = &&\sum_{MM'}  \sum_{l_il_i'm_im_i'}  B_{Ll_1 l_1'}^{XYZ} B_{L'l_2 l_2'}^{XYZ}\left ( \begin{array}{ c c c }
     L & l_1 & l_1' \\
     M & m_1 & m_1'
  \end{array} \right)\left ( \begin{array}{ c c c }
     L' & l_2 & l_2' \\
     M' & m_2 & m_2'
  \end{array} \right) \nonumber \\
&& \times {1\over 6}\big \{ 
[\tilde C^{XX}]_{LM,L'M'} [\tilde C^{YY}]_{l_1m_1,l_1'm_1'}[\tilde C^{XZ} ]_{l_2m_2,l_2'm_2'} + 
[\tilde C^{XY}]_{LM,l_1'm_1'} [\tilde C^{YZ}]_{l_1m_1,L'M'}
[\tilde C^{ZX} ]_{l_2m_2,l_2'm_2'} + \nonumber \\
&& [\tilde C^{XZ}]_{LM,l_2'm_2'} [\tilde C^{ZY}]_{l_2m_2,l_1'm_1'}
[\tilde C^{YX} ]_{l_1m_1,L'M'} + 
 \qtwo [\tilde C^{XX}]_{LM,L'M'} [\tilde C^{ZY}]_{l_2m_2,l_1'm_1'}
[\tilde C^{ZY} ]_{LM,l_2'm_2'} + \nonumber \\
&& [\tilde C^{YY}]_{l_1m_1,l_1'm_1'} [\tilde C^{XZ}]_{LM,l_2'm_2'}[\tilde C^{ZX} ]_{l_2m_2,L'M'} +
[\tilde C^{ZZ}]_{l_2m_2,l_2'm_2'} [\tilde C^{XY}]_{LM,l_1'm_1'}
[\tilde C^{YX}]_{l_1m_1,L'M'} \big \}.\label{MainFisher}
\end{eqnarray}

\n
In the case of near all-sky experiments the off-diagonal elements of the Fisher
matrix will be relatively smaller. The diagonal elements as before 
can be scaled by $f_{sky}$ (the fraction of the sky covered by a near all-sky experiment). The 
covariance matrices can now be expressed only as a function of the related
$C_l$s the auto- and cross-correlation power spectra:

\begin{eqnarray}
F_{LL'} = && \delta_{LL'} \sum_{ll'}\Big \{  B_{Lll'}^{XYZ} B_{Lll'}^{XYZ}
{1 \over \myC_L^{XX}}{1 \over \myC_{l}^{YY}}{1 \over \myC_{l'}^{ZZ}}  + 
B_{Lll'}^{XYZ} B_{Ll'l}^{XYZ}{1 \over \myC_L^{XX}}\left ( {\myC_{l}^{YZ} \over \myC_{l}^{YY} \myC_{l}^{ZZ}}\right )\left ( {\myC_{l'}^{YZ} 
\over \myC_{l'}^{YY} \myC_{l'}^{ZZ}}\right ) \Big \}
\nonumber  \\
&& + \sum_{l} \Big \{ B_{L' l L}^{XYZ} B_{L l L'}^{XYZ}
\left ( {1 \over \myC_l^{YY}}\right )\left ( {\myC_{L'}^{XZ} \over \myC_{L'}^{XX} \myC_{L'}^{ZZ}}\right )\left ( {\myC_L^{ZX} \over \myC_L^{XX} \myC_L^{ZZ}}\right )+ B_{L L' l}^{XYZ} B_{L' L l}^{XYZ}
 \left ( {1 \over \myC_l^{ZZ}}\right )\left ( {\myC_L^{XY} \over \myC_L^{YY} \myC_L^{XX}}\right )\left ( {\myC_{L'}^{YX} \over \myC_{L'}^{XX} \myC_{L'}^{YY}}\right )
 \nonumber \\
&& + B_{LL'l}^{XYZ} B_{l L L'}^{XYZ} \left ( {\myC_L^{XY} \over \myC_L^{XX} \myC_L^{YY}}\right )\left ( {\myC_{L'}^{YZ} \over \myC_{L'}^{YY} \myC_{L'}^{ZZ}}\right )\left ( {\myC_l^{ZX} \over \myC_l^{XX} \myC_l^{ZZ}}\right )+ B_{LlL'}^{XYZ} B_{L'lL}^{XYZ} 
 \left ( {\myC_L^{XZ} \over \myC_L^{XX} \myC_L^{ZZ}}\right )\left ( {\myC_l^{ZY} \over \myC_l^{YY} \myC_l^{ZZ}}\right )\left ( {\myC_{L'}^{YX} \over \myC_{L'}^{XX} \myC_{L'}^{YY}}\right )
\Big \}.
\label{eq:cls_xyz}
\end{eqnarray}
 
\n
The other two terms represented by cyc.perm. consist of terms with suitable
permutations of superscripts $X$,$Y$ and $Z$. In deriving these expressions 
all-sky limits of $\tilde C_{lm,l'm'}^{XX} = ( 1 / \myC_l^{XX} ) \delta_{ll'} \delta_{mm'}$
were used along with the fact that we can write $\tilde C_{lm,l'm'}^{XY} = 
{\{ \myC^{XY}_l / \myC^{XX}_l\myC^{YY}_l \} } \delta_{ll'} \delta_{mm'}$ for all sky case.
For the case when the cross-correlation among two or more fields vanish the expression
simplifies considerably.
   
\n
In case of joint estimation of several bispectra from the same data one can
write the following expression:

\begin{equation}
\hat E_L^{\alpha}[x_i] = \sum_{L'} \sum_{\beta} [N^{-1}]_{LL'}^{\alpha\beta} \left \{ Q_{L'}^\beta[\tilde x_i] - \sum_{i=1,2,3}[\tilde x_i]_{lm} \langle \partial_{lm} Q_{L'}^\beta[\tilde x_i]\rangle_{MC}) \right \}.
\label{eq:xyy_est}
\end{equation}

\n
Here the indices $\alpha$ and $\beta$ correspond to various power spectra $[B^{XYZ}_{l_1l_2l_3}]^\alpha$
or $[B^{XYZ}_{l_1l_2l_3}]^\beta$ which are associated
with bispectra that can be jointly estimated from the same data $[\tilde x_i]$. Below we consider 
two special cases for the skew spectra that we have considered so far.
The expressions for $F^{\alpha\beta}_{LL'}$ and $N^{\alpha\beta}_{LL'}$ can be obtained simply
by replacing the product $B_{Ll_1 l_1'}^{XYZ} B_{L l_1 l_1'}^{XYZ}$ by 
$[B^{XYZ}_{l_1l_2l_3}]^\alpha [B^{XYZ}_{l_1l_2l_3}]^\beta$. In certain situation when accurate noise
modelling is difficult or unlikely an approximate proxy for $C^{-1}$ is used in the form of 
a regularization matrix $R$ which acts as a smoothing of the data. The resulting data vector
$\tilde x_i^R = Rx_i$ is now used for developing a unbiased but suboptimal estimator by replacing 
$\tilde x_i$ with $\tilde x_i^R$.

\begin{figure}
\begin{center}
{\epsfxsize=12. cm \epsfysize=5. cm {\epsfbox[28 507 590 754]{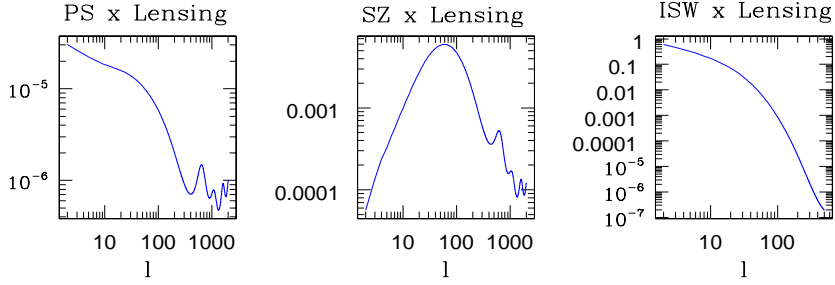}}}
\end{center}
\caption{The cross-spectra $b_l$ introduced in Eq.(\ref{eq:bispec_intro}) 
required for the construction of the bispectra is 
plotted for point source(PS) cross lensing (right panel), ISW cross lensing (middle panel) and SZ cross lensing (left panel) bispectrum as a function of $l$. See text for details.}
\label{fig:bicls}
\end{figure}

\subsubsection{Special Case (A): Z=Y}

\n
The estimator in this case corresponds to $E_l^{X,Y^2}$.

\begin{eqnarray}
&& F_{LL'} = 2 \delta_{LL'}\sum_{ll'}\Big \{B_{Lll'}^{XYY} B_{Lll'}^{XYY}
 {1 \over \myC_L^{XX}}{ 1 \over \myC_l^{YY}}{1 \over \myC_l^{YY}} \Big \}  + \sum_{l} \Big \{
 2~B_{LL'l}^{XYY} B_{L'Ll}^{XYY}\left (  { \myC_L^{XY} \over \myC_L^{XX} \myC_L^{YY}} \right ) \left ( { \myC_L^{XY} \over \myC_L^{XX} \myC_L^{YY}} \right ){1 \over \myC_l^{YY}} 
\nonumber \\ 
&& +B_{LL'l}^{XYY} B_{lLL'}^{XYY}  
{1 \over \myC_{L'}^{YY}}\left ( {\myC_L^{XY} \over \myC_L^{XX} \myC_L^{YY}} \right )\left ( {\myC_l^{YX} \over \myC_l^{XX} \myC_l^{YY}} \right )  + 
B_{LlL'}^{XYY} B_{L'lL}^{XYY}
{1 \over \myC_l^{YY}}\left ( {\myC_L^{XY} \over \myC_L^{XX} \myC_L^{YY}} \right )\left ( {\myC_{L'}^{YX} \over \myC_{L'}^{XX} \myC_{L'}^{YY}} \right )  \Big \}.
\label{eq:xyy_fish}
\end{eqnarray}

\subsubsection{Special Case (B): Z=Y=X}

\begin{eqnarray}
&& F_{LL'} = 2 \delta_{LL'}\sum_{ll'}\Big \{B_{Lll'}^{XXX} B_{Lll'}^{XXX}
{1 \over \myC_L^{XX}}{ 1 \over \myC_l^{XX}}{1 \over \myC_l^{XX}} \Big \} + 
\sum_{l} \Big \{ B_{LL'l}^{XXX} B_{LL'l}^{XXX} {1 \over \myC_L^{XX}}{ 1 \over \myC_{L'}^{XX}}{1 \over \myC_l^{XX}} \Big \}
\end{eqnarray}

\subsubsection{Special Case (C): Z=X} 

\n
The estimator we consider here is $E_l^{X,XY}$ can not derived by simple identification of superscript.
We define the new functions related to the estimators $Q_L$,   $\partial_{lm}^X \hat Q_L$ and
$\partial_{lm}^Y \hat Q_L$ according to the same prescription above.

\begin{eqnarray}
&& \hat Q_L[\tilde X,\tilde Y] \equiv \sum_{M} \tilde X_{LM} \sum_{l'm',l''m''} B_{Ll'l''}\left ( \begin{array}{ c c c }
     L & l' & l'' \\
     M & m' & m''
  \end{array} \right) \tilde X_{l'm'} \tilde Y_{l''m''} \\
&& \partial_{lm}^Y \hat Q_L[\tilde X,\tilde Y] \equiv \sum_M  \tilde X_{LM} \sum_{l'm',l''m''} B_{Ll'l''}\left ( \begin{array}{ c c c }
     L & l' & l'' \\
     m & m' & m''
  \end{array} \right) \tilde X_{l'm'} \\
&&  \partial_{lm}^X \hat Q_L[\tilde X,\tilde Y] \equiv   \sum_{LM} \tilde X_{LM}
\sum_{l'm'} B_{Lll'}\left ( \begin{array}{ c c c }
     L & l & l' \\
     M & m & m'
  \end{array} \right) \tilde X_{l'm'} + \delta_{Ll}\delta_{Mm} 
\sum_{l'm'} B_{Lll'}\left ( \begin{array}{ c c c }
     L & l & l' \\
     M & m & m'
  \end{array} \right) \tilde X_{l'm'}\tilde Y_{l''m''}.
\label{eq:q_dq_xyx}
\end{eqnarray}

\n
The estimator in this case takes the form:

\begin{equation}
\hat E_L[x_i] =  [N^{-1}]_{LL'} \left \{ Q_{L'}[\tilde x_i] - \sum_{i=1,2}[\tilde x_i]_{lm} \langle \partial^i_{lm} Q_{L'} [\tilde x_i]\rangle_{MC}) \right \}
\label{eq:est_xyx}
\end{equation}

\begin{eqnarray}
&& F_{LL'} = \delta_{LL'} \sum_{ll'}\Big \{ B_{Lll'}^{XYX} B_{Lll'}^{XYX}{1 \over \myC_L^{XX}}{ 1 \over \myC_{l}^{XX}}{1 \over \myC_{l'}^{XX}} \Big \} 
+  \Big \{ B_{Lll'}^{XYX} B_{Lll'}^{XYX} {1 \over \myC_L^{XX}} \left ( { \myC_l^{XY} \over \myC_{l}^{XX} \myC_l^{YY}} \right ) \left ( {\myC_{l'}^{XY} \over \myC_{l'}^{XX} \myC_{l'}^{YY}}  \Big \} \right )\\
 && + \sum_{l}\Big \{ B_{LL'l}^{XYX} B_{L'Ll}^{XYX}{\left ( \myC_L^{XY}\over \myC_L^{XX}\myC_L^{YY} \right )}{ \left ( \myC_{L'}^{XY} \over \myC_{L'}^{XX} \myC_{L'}^{YY}\right )}{1 \over \myC_{l}^{XX}} 
 +  B_{LlL'}^{XYX} B_{L'lL}^{XYX}{\myC_L^{XY}\over \myC_L^{XX}\myC_L^{YY}}{ \myC_{l}^{XY} \over \myC_{l}^{XX} \myC_{l}^{YY}}{1 \over \myC_{L'}^{XX}} \nonumber \\
&& +  B_{LL'l}^{XYX} B_{L'lL}^{XYX}{ 1 \over \myC_L^{XX}}{\left (  \myC_{L'}^{YX} \over \myC_{L'}^{XX} \myC_{L'}^{YY} \right )}{\left ( \myC_l^{XY} \over \myC_{l}^{XY} \myC_l^{XY} \right )} 
+  B_{LlL'}^{XYX} B_{L'lL}^{XYX}{1 \over \myC_L^{XX}}{ 1 \over \myC_{L'}^{XX}}{1 \over \myC_{l}^{YY}} \Big \}
\label{eq:xyx_fish}
\end{eqnarray}

\section{Specific Examples}
\label{sec:example}

\n
The discussion so far has been completely general. We specialise now for a few practical cases of 
cosmological importance. These correspond to the study of mixed bispectra associated  with lensing 
induced correlation of secondaries and CMB as well as frequency cleaned SZ catalogs against CMB sky.

\subsection{Lensing Reconstruction}

\subsubsection{One-point estimator:}

\n
Various estimators  associated with lensing reconstruction were introduced by different authors,
e.g.\citep{Hu00,huoka02}). It was recently studied by \citet{SmZaDo00} and was used to 
probe effect of lensing in CMB by cross-correlating with external data-set
such as NVSS survey against WMAP observations.

\begin{equation}
S^{lens} = {1 \over 2 N} \sum_{l_i m_i} B_{l_1l_2l_3}^{\hhs}
\left [ \tilde \delta_{l_1m_1}\tilde \delta_{l_2m_2}\tilde \psi_{l_3m_3} - [C^{\phi\phi}]^{-1}_{l_1m_1,l_2m_2}\tilde \psi_{l_3m_3}\right ].
\end{equation} 

\n
This is achieved by writing the reconstructed lensing potential in terms of the CMB harmonics
and cross-correlating it with low-redshift large-scale tracers such as galaxy surveys \citep{SmZaDo00}.
The bispectrum $B_{l_1l_2l_3}^{\hhs}$ depends in addition to the $C_l$s of the CMB multipole,
on the cross-correlation between the CMB sky $\delta(\oh)$ and the low-redshift tracer
field $\psi(\oh)$.
The reduced bispectrum of interest $b^{\hhs}_{l_1l_2l_3}$ and the related form factor $f_{l_1l_2l_3}$
can be written as:

\begin{eqnarray}
&& b^{\hhs}_{l_1l_2l_3}= \Big \{ f_{l_1l_2l_3}\myC_{l_2}^{\delta\delta}+f_{l_2l_1l_3}\myC_{l_1}^{\delta\delta} \Big \}\myC_{l_3}^{\delta\psi} \nonumber \\
&& f_{l_1l_2l_3}= {1 \over 2} \Big \{ l_2(l_2+1) +l_3(l_3+1)-l_1(l_1+1) \Big \}
\end{eqnarray}

\n
The multipole  $\delta_{l_1m_1}$ and $\delta_{l_2m_2}$ are
associated with the CMB sky and
$\psi_{l_3m_3}$ is the multipole associated with the large-scale structure tracer
at low  redshift and hence correlates with the lensing potential 
(e.g. NVSS survey). The above estimator directly probes the cross-correlation
between the lensing potential harmonics $\phi_{lm}$ constructed from temperature 
harmonics $\delta_{lm}$ and the harmonics of the tracers $\psi_{lm}$.
It is interesting to notice that the estimator 
constructed lacks the term which signifies the correlation between
$\delta(\oh)$ and $\psi(\oh)$ through the coupling $C^{\delta\psi}$. 
Though the bispectrum itself depends
directly on the cross-power spectra. Using the results derived 
before we can write the Fisher matrix associated  with this estimator 
can be written as:

\begin{equation}
F = N^{-1}= {1 \over 2 }\sum_{l_im_i} B_{l_1l_2l_3}^{\hhs}B_{l_4l_5l_6}^{\hhs} { [C^{\delta\delta}]}^{-1}_{l_1m_1,l_4m_4}{[C^{\delta\delta}]}^{-1}_{l_2m_2,l_5m_5}
{[C^{\psi\psi}]}^{-1}_{l_3m_3,l_6m_6}.
\end{equation}

\subsubsection{Estimators for the skew spectrum}

\n
If instead of the one-point estimator described above, we compute the two-point estimator or the skew spectrum as follows:

\begin{equation}
E_L[\tilde \delta, \tilde \psi] =[N^{-1}]_{LL'}\left \{ Q_{L'}[\tilde \delta, \tilde \psi] - 
\tilde \psi \langle \partial_{lm}^{\tilde \psi} Q_{L'}[\tilde \delta,\tilde \psi] \rangle \right \}.
\end{equation}

\n
The corresponding expressions for the functions $Q_{L'}[\tilde \phi]$ and 
$\partial_{lm} Q_{L'}[\tilde \phi,\tilde \psi]$ are given by: 

\begin{equation}
Q_{L} = \sum_{M} \tilde \psi_{LM} \sum_{lm,lm} \tilde \delta_{lm} \tilde \delta_{l'm'} \left ( \begin{array}{ c c c }
     L & l & l' \\
     m & m & m'
  \end{array} \right);
\qtwo
\partial_{lm}^{\psi} Q_{L} =  \sum_{LM} {\psi}_{LM} \sum_{l'm'} B_{Lll'}^{\hhs} {\phi}_{l'm'}\left ( \begin{array}{ c c c }
     L & l & l' \\
     m & m & m'
  \end{array} \right).
\label{eq:lens_der}
\end{equation}

\n
Corresponding Fisher matrices can which turns out to be diagonal can be written as:

\begin{equation}
F_{LL'} = N^{-1}_{LL'} =
\left \{ \langle \partial_{lm}^{\psi} Q_{L} [C^{\psi\psi}]_{lm,l'm'} \partial_{l'm'}^\psi Q_{L'} \rangle
-\langle \partial_{lm}^{\psi} Q_{L} \rangle [C^{\psi\psi}]_{lm,l'm'}\langle \partial_{l'm'}^\psi Q_{L'} \rangle \right \};
\label{eq:lens_fish}
\end{equation}

\n
which finally leads us to:

\begin{equation}
F_{LL'}=  {1 \over 2} \sum_{ll'} B^{\hhs}_{l_1l_2L}B^{\hhs}_{l_3l_4L'}{ [C^{\phi\phi}]}^{-1}_{l_1m_1,l_3m_3}
{ [C^{\phi\phi}]}^{-1}_{l_2m_2,l_4m_4}{ [C^{\psi\psi}]}^{-1}_{LM,L'M'}.
\end{equation}

\n
In the limit of all sky survey and homogeneous noise we can write:

\begin{equation}
F_{LL'}=  \delta_{LL'} \sum_{ll'} B^{\hhs}_{ll'L}B^{\hhs}_{ll'L'}
{\left ( 1 \over \myc_L^{\psi\psi} \right )} {\left (1 \over \myC_l^{\phi\phi} \right )} 
{\left ( 1 \over \myC_{l'}^{\psi\psi} \right )} 
\end{equation}

\n
A comparison with the previous estimator shows the presence of off-diagonal
entries in the Fisher matrix even if direct correlation between ${\delta}$
and $\psi$ are absent in the estimator.

It is possible to work with CMB sky without external data sets
(such as NVSS or other galaxy surveys) to probe weak lensing e.g. the
power spectrum of the lensing potential itself 
is related to four-point statistics of the temperature - which
makes it noise dominated. use of external tracers such as galaxy surveys
can reduce the problem to three-point level thus lowering the need on
sensitivity of the instrument. The discussion above can have direct relevance 
for use of other tracers such as the one with neutral hydrogen observations \citep{ZZ06}.

\begin{figure}
\begin{center}
{\epsfxsize=12. cm \epsfysize=5. cm {\epsfbox[28 507 590 754]{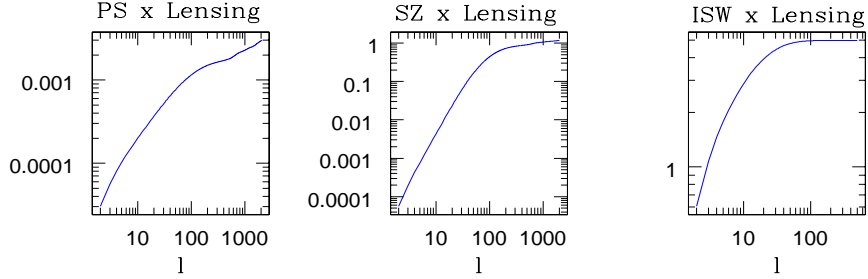}}}
\end{center}
\caption{The cross-spectra $b_l$ introduced in Eq.(\ref{eq:bispec_intro}) 
required for the construction of the bispectra is 
plotted for ISW cross lensing (right panel) and SZ cross lensing (left panel) bispectrum as a function of $l$. See text for details.}
\label{fig:bicls4}
\end{figure}

\subsection{Sunyaev-Zeldovich-$\rm CMB^2$ mixed bispectrum}

The secondary bispectrum caused by the Sunyaev-Zeldovich effect is one of the 
most pronounced secondary bispectrum among many others 
\citep{GoldbergSpergel99,SperGold99,CoorayHu,cooSZ2,Cooray1,Cooray2}. 
Following \citet{CoorayHuTeg} we study if frequency cleaned maps of all-sky CMB and SZ maps can also 
be used to construct power-spectra associated with the mixed bispectra
with signal-to-noise ratio that can be detectable with ongoing CMB 
experiments. It probes mode-coupling effects generated by correlation involved 
with gravitational lensing angular deflections in CMB and the SZ
effects due to large-scale pressure fluctuations. As before the 
estimator which can be constructed from the CMB $\tilde a_{lm}$ and 
the Sunyaev-Zeldovich $s_{lm}$ multipoles. There is a possibility
of constructing the correlating the product map $a(\oh)s(\oh)$ with $a(\oh)$
as well as $a(\oh)$ and $s(\oh)^2$. In terms of the suboptimal estimators
introduced before the correspond to $C_l^{as,a}$ and $C_l^{s,a^2}$
respectively. In the second case analysis is exactly same as that of
lensing reconstruction discussed before. However in the first case the 
optimal estimator is expressed as follows:

\begin{equation}
E_L^{s,aa}[\tilde a, \tilde s] =[N^{-1}]_{LL'}\left \{ Q_{L'}[\tilde a,\tilde s] - 
\tilde a_{lm} \langle \partial_{lm}^a Q_{L'}[\tilde a,\tilde s] \rangle 
-\tilde s_{lm} \langle \partial_{lm}^s Q_{L'}[\tilde a,\tilde s] \rangle \right \}.
\end{equation}

\n
The above estimator considering is same as Eq.(\ref{eq:xyy_est});
with the corresponding $Q_{L'}[\tilde a,\tilde s]$ function and its derivatives
are given in Eq.(\ref{eq:xyy_qdq}).

\n
The mixed bispectrum of $\rm CMB^2-SZ$ is known to be exactly same as
that of bispectrum we considered in  the lensing reconstruction.
This is true not only for SZ-lensing bispectrum but for 
other lensing-induced correlation-related bispectra too.
The only difference is in different $b_l$s involved.

\begin{eqnarray}
&& F_{LL'} = 2 \delta_{LL'}\sum_{ll'}\Big \{B_{Lll'}^{saa} B_{Lll'}^{saa}
 {1 \over \myc_L^{ss}}{ 1 \over \myc_l^{aa}}{1 \over \myc_l^{a}} \Big \}  + \sum_{l} \Big \{
 2~B_{LL'l}^{saa} B_{L'Ll}^{saa}\left (  { \myc_L^{sa} \over \myc_L^{ss} \myc_L^{aa}} \right ) \left ( { \myc_L^{sa} \over \myc_L^{ss} \myc_L^{aa}} \right ){1 \over \myc_l^{aa}} 
\nonumber \\ 
&& +B_{LL'l}^{saa} B_{lLL'}^{saa}  
{1 \over \myc_{L'}^{aa}}\left ( {\myc_L^{sa} \over \myc_L^{ss} \myc_L^{aa}} \right )\left ( {\myc_l^{sa} \over \myc_l^{ss} \myc_l^{aa}} \right )  + 
B_{LlL'}^{saa} B_{L'lL}^{saa}
{1 \over \myc_l^{aa}}\left ( {\myc_L^{sa} \over \myc_L^{ss} \myc_L^{aa}} \right )\left ( {\myc_{L'}^{sa} \over \myc_{L'}^{ss} \myc_{L'}^{aa}} \right )  \Big \}.
\end{eqnarray}

\n
This is an application of the case considered in Eq.~(\ref{eq:xyy_fish}).
For the one-point mixed skewness associated with this power spectra the 
related Fisher error is simply given by the sum over all the elements.
$F = \sum_{L,L'} F_{LL'}$. The cross correlations $\myc_l^{sa}$ between the two maps
$s(\oh)$ and $a(\oh)$ introduces the off-diagonal elements in the Fisher matrix 
even for the case of 
all-sky coverage and homogeneous noise. Ignoring the correlations we can
recover the Fisher matrix elements derived for the case of lensing reconstruction.

In addition to considering the cross-correlation of $s(\oh)$ and $a^2(\oh)$
as discussed above, the other estimator of non-Gaussianity that we can consider is
by considering the cross-correlation of product field $s(\oh)a(\oh)$ and $a(\oh)$ 
which is same as the estimator is same as that defined in Eq.~(\ref{eq:est_xyx})
with the relevant $Q$ term and its derivative given by Eq.~(\ref{eq:q_dq_xyx}). The 
associated Fisher matrix is given by Eq.~(\ref{eq:xyx_fish}). The one-point estimator
recovered from both of these degenerate estimators will be the same.

\section{Conclusions}
\label{sec:conclu}

Extending previous work for estimation of power-spectra from correlated data sets 
we show how pseudo-$C_l$-based approaches (PCL) can be used for estimation of cross-correlation
power spectra from multiple cosmological surveys through a joint analysis.
Analytical results were derived under very general conditions using an arbitrary
mask as well as arbitrary noise properties. We also keep the weighting of the 
data completely general. Our analytical results also include a systematic 
analysis of covariance of various deconvolved $\hat C_l$s characterizing 
auto- and cross-power spectra from a joint analysis. While PCL-based approaches
are known to be unbiased they are not in general optimal. However they
can be made to act in a near-optimal way by the introduction of weights in
different regimes corresponding to signal or noise domination. These studies 
will be useful in analyzing simulated as well as real survey data either in
projection or in 3D. We specialise these expressions to recover well-known 
$f_{sky}$ approximation used in the literature for the error analysis.
Using a halo model inspired approach we compute the expected cross-correlation
signal in cross-correlating NVSS type survey with the CMB sky through the
ISW effect. We also study the cross-correlation between the frequency-cleaned
SZ surveys against the ISW effect. The cross-correlation study also provides
the covariances among different estimated $C_l$s and the signal-to-noise 
of detection for a specific survey. However we want to stress that the formalism
developed here is more powerful and can tackle many issues
in analysing realistic surveys. A detailed study using simulations will
be presented elsewhere.

The analysis of the bispectrum is one step beyond the power-spectrum and provides
additional cosmological information. The primary
motivation to date has been to put constraints on early-universe scenarios,
however secondary bispectra can play a significant role
in enhancing our understanding of large-scale structure formation scenarios.
The secondary bispectrum is mainly related to mode-coupling by secondary effects and
lensing. We study various statistics which can directly
handle realistic data sets. 
Extending previous work by \cite{MuHe09}, we take into account multiple
correlated fields which are used for constructing a mixed skewness at one-point level
as well as constructing a skew spectrum at the level of two-point.
A very general framework was developed for the study of bispectrum from 
correlated fields in an unbiased and optimized way. We introduce the inverse
covariance weighting and specialize our results for the analysis of 
bispectrum originating from lensing-secondary correlations.
A simple-minded approach which handles the noise and partial sky coverage 
in a nearly optimal way using Monte Carlo techniques is also
discussed. We also develop an approach based on PCL to study
the skew spectrum. This approach, whilst 
suboptimal, can handle the noise and partial sky coverage directly.
It is also possible to use weights to make it near optimal in
the limit of high $l$,  and can be useful mainly because of its speed of handling MC realisations.  In its most general form, the estimator $E_L$ (equation (\ref{MainEstimator}) for the skew spectrum of mixed fields includes the effect of partial sky coverage and inhomogeneous noise, and provides a compact function which can be compared with theoretical models to identify the source of the correlations between the fields.  The associated Fisher matrix (equation \ref{MainFisher}) allows statistical analysis of the the $E_L$ estimates, allowing the estimation of the relative contributions from different physical processes.

For specific examples we have focussed on probing the secondary non-Gaussianity
with Planck type all-sky experiments and surveys such as NVSS. 
The signal to noise ratios for cross-correlation studies involving
lensing potential and secondaries such as SZ and ISW would allow detection
with Planck. However to differentiate among various effects one would
need to go beyond cumulative signal-to-noise estimates and the statistics
which we introduced here will be useful diagnostic tools.

There has been lot of work by a number of authors to detect correlations between 
the WMAP CMB and large scale structures, which typically conclude with a constraint 
on the dark energy (accelerating universe). Analysis of secondary bispectrum has also
been attempted. However, at this point, consistent simulations which can correctly take into
account, the correlation between CMB and the LSS, and the impact of the LSS on the various 
observables is still remains to be developed. Though a patchwork of simulations 
are getting ready, we still lack suitable simulations which can be used
both for cross-correlational analysis or the entire range of bispectrum analysis
at the moment. Our approach can be invaluable in quantifying accuracy of
such consistency check and eventually to put constrain on cosmology using real
high resolution data. We have not taken into account the errors or residuals 
from foreground removals. Some of the foreground contaminations may well be
correlated to various LSS tracers. These issues and how PCL based approach can
tackle them will be dealt with elsewhere.

\section{Acknowledgements}
\label{acknow}

DM was supported by a STFC rolling grant at Royal Observatory at Edinburgh, Institute 
for Astronomy, when this work was performed. It is a pleasure to acknowledge useful 
exchanges with Jacques Delabrouille, Matthias Bartelmann and Patricio Vielva Martinez.
AC acknowledges support from NSF CAREER AST-0645427.

{}


\begin{thebibliography}{}


\bibitem[\protect\citeauthoryear{Acquaviva et al.}{2003}]{Acq03}
Acquaviva V., Bartolo N., Matarrese S., Riotto A., 2003, Nucl. Phys. B667, 119

\bibitem[\protect\citeauthoryear{Alishahiha, Silverstein \& Tong}{2004}]{Ali04}
Alishahiha M., Silverstein E., Tong T., 2004, Phys. Rev. D70, 123505

\bibitem[\protect\citeauthoryear{Arkani-Hamed et al.}{2004}]{Ark04}
Arkani-Hamed N., Creminelli P., Mukohyama S., Zaldarriaga M., 2004, JCAP0404:001

\bibitem[\protect\citeauthoryear{Afshordi,Loh \& Strauss}{2004}]{Afs1} 
Afshrodi N., Loh, Y., Strauss, M.A., 2004, Phys. Rev. D 69, 083524

\bibitem[\protect\citeauthoryear{Afshordi}{2004}]{Afs2} 
Afshrodi N., 2004, Phys. Rev. D, 70, 083536

\bibitem[\protect\citeauthoryear{Babich}{2005}]{Babich}
Babich D.,  2005, Phys. Rev. D72, 043003

\bibitem[\protect\citeauthoryear{Babich \& Pierpaoli}{2008}]{Babich08} 
Babich D., Pierpaoli E., 2008, Phys. Rev. D77, 123011 

\bibitem[\protect\citeauthoryear{Babich \& Zaldarriaga}{2004}]{BaZa04} 
Babich D \& Zaldarriaga M., 2004, Phys. Rev. D70, 083005

\bibitem[\protect\citeauthoryear{Babich, Creminelli \& Zaldarriaga}{2004}]{BCP04} 
Babich D., Creminelli P., Zaldarriaga M., 2004, JCAP, 8, 9 

\bibitem[\protect\citeauthoryear{Bartolo, Matarrese \& Riotto}{2006}]{Bartolo06}
Bartolo N., Matarrese S., Riotto A., 2006, JCAP, 06, 024

\bibitem[\protect\citeauthoryear{Buchbinder, Khoury \& Ovrut}{2008}]{Buch08}
Buchbinder E.I., Khoury J., Ovrut B.A., 2008, Phys.Rev.Lett,100:171302

\bibitem[\protect\citeauthoryear{Boughn \& Crittenden}{2005}]{BoughCrit1} 
Boughn S., Crittenden R., MNRAS, 2005, 360, 1013

\bibitem[\protect\citeauthoryear{Boughn \& Crittenden}{2004a}]{BoughCrit2} 
Boughn S., Crittenden R., Nature 427 (2004) 45

\bibitem[\protect\citeauthoryear{Boughn \& Crittenden}{2004b}]{BoughCrit3} 
Boughn S., Crittenden R., New Astron.Rev. 49 (2005) 75-7

\bibitem[\protect\citeauthoryear{Cabella et al. }{2006}]{Cabella06}
Cabella P., Hansen F.K., Liguori M., Marinucci D., Matarrese S., Moscardini L., Vittorio N., 2006, MNRAS, 369, 819

\bibitem[\protect\citeauthoryear{Cabre et al.}{2006}]{Cab}
Cabre A., Gaztanaga E., Manera M., Fosalba P., Castander F.
Mon.Not.Roy.Astron.Soc.Lett., 372 (2006) L23-L27

\bibitem[\protect\citeauthoryear{Castro}{2004}]{Castro04}
Castro P., 2004, Phys. Rev. D67, 044039 (erratum D70, 049902)

\bibitem[\protect\citeauthoryear{Chen, Huang \& Kachru}{2006}]{Chen06}
Chen X., Huang M., Kachru S., Shiu G., 2006, hep-th/0605045

\bibitem[\protect\citeauthoryear{Chen, Easther \& Lim}{2007}]{Chen07}
Chen X., Easther R., Lim E.A., 2007, JCAP, 0706:023

\bibitem[\protect\citeauthoryear{Chen \& Szapudi }{2006}]{ChSz06}
Chen G. \& Szapudi I., Astrophys.J, 647, 2006, L87-L90, 2006

\bibitem[\protect\citeauthoryear{Cheung et al.}{2008}]{Cheung08}
Cheung C., Creminelli P., Fitzpatrick A.L., Kaplan J., Senatore L., 2008, JHEP, 0803, 014

\bibitem[\protect\citeauthoryear{Cooray \& Hu}{2000}]{CoorayHu}
Cooray A.R., Hu W., 2000, ApJ, 534, 533-550

\bibitem[\protect\citeauthoryear{Cooray, Hu \& Tegmark}{2000}]{CoorayHuTeg}
Cooray A., Hu W., Tegmark, M. 2000, ApJ, 540, 1-13

\bibitem[\protect\citeauthoryear{Cooray}{2001a}]{Cooray1} 
Cooray A., 2001a, Phys. Rev. D, 64, 043516 

\bibitem[\protect\citeauthoryear{Cooray}{2001b}]{Cooray2} 
Cooray A., 2001b, Phys Rev. D, 64, 063514

\bibitem[\protect\citeauthoryear{Cooray}{2006}]{Cooray3} 
Cooray A., 2006, Phys. Rev. Lett., 97, 261301

\bibitem[\protect\citeauthoryear{Cooray, Li \& Melchiorri}{2008}]{Cooray8}
Cooray A., Li C., Melchiorri A., 2008, Phys. Rev. D77,103506

\bibitem[\protect\citeauthoryear{Cooray}{2000}]{cooSZ2} 
Cooray A., 2000, Phys. Rev. D, 62, 103506 

\bibitem[\protect\citeauthoryear{Cooray \& Seth}{2000}]{sethcoo}
Cooray A. Seth R., Phys. Rept. 372 (2002) 1-129

\bibitem[\protect\citeauthoryear{Creminelli}{2003}]{Crem03}
Creminelli P., 2003, JCAP 0310, 003

\bibitem[\protect\citeauthoryear{Creminelli et al.}{2006}]{Crem06} 
Creminelli P., Nicolis A., Senatore L., Tegmark M., Zaldarriaga M., 2006, JCAP, 5, 4 

\bibitem[\protect\citeauthoryear{Creminelli et 
al.}{2007}]{Crem07a} Creminelli P., Senatore L., Zaldarriaga M., Tegmark M., 2007, JCAP, 3, 5 

\bibitem[\protect\citeauthoryear{Creminelli, Senatore, 
\& Zaldarriaga}{2007}]{Crem07b} Creminelli P., Senatore L., Zaldarriaga M., 2007, JCAP, 3, 19 

\bibitem[\protect\citeauthoryear{Corsanti, Giannantonio, Melchiorri}{2004}]{Cor1} 
Corasaniti, P.S., Giannantonio T., Melchiorri, A.,  2005, Phys, Rev D 72, 023514 

\bibitem[\protect\citeauthoryear{Diego, Silk, Silwa}{2004}]{Die1} 
Diego,J.M., Silk J., Sliwa W.,  New Astron.Rev. 47 (2003a) 855, 

\bibitem[\protect\citeauthoryear{Diego,Silk, Silwa}{2004}]{Die2} 
Diego,J.M., Silk J., Sliwa W., Mon.Not.Roy.Astron.Soc. 346 (2003b) 940

\bibitem[\protect\citeauthoryear{Efstathiou}{2004}]{Efs1} 
Efstathiou G., 2004, MNRAS, 349, 603 

\bibitem[\protect\citeauthoryear{Efstathiou}{2006}]{Efs2} 
Efstathiou G., 2006, MNRAS, 370, 343 

\bibitem[\protect\citeauthoryear{Falk et al.}{1993}]{Falk93}
Falk T., Madden R., Olive K.A., Srednicki M., 1993, Phys. Lett. B318, 354

\bibitem[\protect\citeauthoryear{Fosabala \& Gaztanaga}{2004}]{FosGaz1} 
Fosabala, P. \& Gaztanaga E., (2004), MNRAS, 350, L37

\bibitem[\protect\citeauthoryear{Fosabala \& Gaztanaga}{2006}]{FosGaz2} 
Fosabala, P. \& Gaztanaga E., Castander F., (2003), ApJ, 597, L89  

\bibitem[\protect\citeauthoryear{Gangui et al.}{1994}]{Gangui94}
Gangui A., Lucchin F., Matarrese S., Mollerach S., 1994, ApJ, 430, 447

\bibitem[\protect\citeauthoryear{Giannantonio et al.}{2002}]{Gian} 
Giannantonio, T. et al.,Phys.Rev. D74 (2006) 063520

\bibitem[\protect\citeauthoryear{Goldberg \& Spergel}{1999}]{GoldbergSpergel99}
Goldberg D.M., Spergel D.N., 1999, Phys. Rev. D59, 103002

\bibitem[\protect\citeauthoryear{Gupta, Berera \& Heavens}{2002}]{GuBeHea02} 
Gupta S., Berera A., Heavens A.F., Matarrese S., 2002, Phys.Rev. D66, 043510

\bibitem[\protect\citeauthoryear{Heavens}{1998}]{Heav98}
Heavens A.F.,  1998, MNRAS, 299, 805

\bibitem[\protect\citeauthoryear{Hivon et al.}{2002}]{Hiv} 
Hivon E., G{\'o}rski K.~M., Netterfield C.~B., Crill B.~P., Prunet S., 
Hansen F., 2002, ApJ, 567, 2 

\bibitem[\protect\citeauthoryear{Hirata et al.}{2008}]{Hi08} 
Hirata C.M., Ho S., Padmanabhan N., Seljak U., Bahcall N., 2008, Phys.Rev.D78, 043520

\bibitem[\protect\citeauthoryear{Ho et. al.}{2008}]{Ho08}
Ho S., Hirata ~C.~M., Padmanabhan N., Seljak U., Bahcall N.,2008, Phys.Rev.D,78, 043519

\bibitem[\protect\citeauthoryear{Hu}{2000}]{Hu00} 
Hu W., 2000, PhRvD, 62, 043007 

\bibitem[\protect\citeauthoryear{Hu \& Okamoto}{2002}]{huoka02} 
Hu W., Okamoto T., 2002, ApJ, 574, 566 

\bibitem[\protect\citeauthoryear{Jaffe \& Kamionkowski}{2004b}]{JaffeKamion} 
Jaffe A.H., Kamionkowski M., Phys.Rev. D58 (1998) 043001

\bibitem[\protect\citeauthoryear{Komatsu \& Spergel}{2001}]{KomSpe01}
Komatsu E., Spergel D.~N., 2001, Phys. Rev. D63, 3002

\bibitem[\protect\citeauthoryear{Komatsu, Spergel \& Wandelt}{2005}]{KSW}
Komatsu E., Spergel D.~N., Wandelt B.~D., 2005, ApJ, 634, 14 

\bibitem[\protect\citeauthoryear{Komatsu et al.}{2002}]{Komatsu02} 
Komatsu E., Wandelt B.~D., Spergel D.~N.,Banday A.~J., G{\'o}rski K.~M., 2002, ApJ, 566, 19 

\bibitem[\protect\citeauthoryear{Komatsu et al.}{2003}]{Komatsu03} 
Komatsu E., et al., 2003, ApJS, 148, 119

\bibitem[\protect\citeauthoryear{Koyama et al.}{2007}]{Koya07}
Koyama K., Mizuno S., Vernizzi F., Wands D., 2007, JCAP 0711:024



\bibitem[\protect\citeauthoryear{Linde \& Mukhanov}{1997}]{lindemukha}
Linde A.~D., Mukhanov V.~F., (1997), Phys.\ Rev.\  D {\bf 56}, 535 

\bibitem[\protect\citeauthoryear{Lyth, Ungarelli \& Wands}{2003}]{Lyth03}
Lyth D.H., Ungarelli C., Wands D., 2003, Phys. Rev. D67, 023503

\bibitem[\protect\citeauthoryear{Maldacena}{2003}]{Mal03}
Maldacena J.M., 2003, JHEP, 05, 013

\bibitem[\protect\citeauthoryear{Medeiros \& Contaldo}{2006}]{MedeirosContaldi06}
Medeiros J., Contaldi C.R, 2006, MNRAS, 367, 39

\bibitem[\protect\citeauthoryear{Moss \& Xiong}{2007}]{Moss}
Moss I., Xiong C., 2007, JCAP, 0704, 007

\bibitem[\protect\citeauthoryear{Munshi, Souradeep \& Starobinsky}{1995}]{MuSoSt95}
Munshi D., Souradeep, T., Starobinsky, Alexei A., 1995, ApJ, 454, 552

\bibitem[\protect\citeauthoryear{Munshi \& Heavens}{2009}]{MuHe09}
Munshi D., Heavens A., arXiv:0904.4478 

\bibitem[\protect\citeauthoryear{Munshi, Melott \& Coles}{2000}]{Mun}
Munshi D., Melott A.L., Coles P., {\it MNRAS}, 2000, 311, 149.

\bibitem[\protect\citeauthoryear{Nolta et. al.}{2004}]{No08}
Nolta et al., Astrophys.J., 2004, 608, 10

\bibitem[\protect\citeauthoryear{Padmanabhan et al.}{2005}]{pad} 
Padmanabhan N., Hirata C. M., Seljak U., Schlegel D., Brinkmann J., Schneider D.P.
Phys.Rev. D72 (2005) 043525

\bibitem[\protect\citeauthoryear{Peiris \& Spergel}{2000}]{Peir} 
Peiris H. V. \& Spergel D.N., 2000, ApJ, 540, 605

\bibitem[\protect\citeauthoryear{Sachs \& Wolfe}{1967}]{SW}
Sachs R. K. \& Wolfe A.M., 1967, ApJ, 147, 73

\bibitem[\protect\citeauthoryear{Salopek \& Bond}{1990}]{Salopek90} 
Salopek D.~S., Bond J.~R., 1990, PhRvD, 42, 3936 

\bibitem[\protect\citeauthoryear{Salopek \& Bond}{1991}]{Salopek91} 
Salopek D.~S., Bond J.~R., 1991, PhRvD, 43, 1005 

\bibitem[\protect\citeauthoryear{Santos et al.}{2003}]{Santos}
Santos M.G. et al., 2003, MNRAS, 341, 623

\bibitem[\protect\citeauthoryear{Serra \& Cooray}{2008}]{SerCoo08} 
Serra P., Cooray A., 2008, Phys. Rev. D, 77, 107305

\bibitem[\protect\citeauthoryear{Smith, Zahn \& Dore}{2007}]{SmZaDo00}  %
Smith K.M., Zahn O., Dore O., 2007, Phys. Rev. D, 76, 043510

\bibitem[\protect\citeauthoryear{Smith \& Zaldarriaga}{2006}]{SmZa06} 
Smith K.~M., Zaldarriaga M., 2006, arXiv:astro-ph/0612571

\bibitem[\protect\citeauthoryear{Smith, Senatore \& Zaldarriaga}{2009}]{SmSeZa09} 
Smith K.M., Senatore L., Zaldarriaga M., 2009,  arXiv:0901.2572

\bibitem[\protect\citeauthoryear{Spergel \& Goldberg}{1999}]{SperGold99}
Spergel D.N., Goldberg D. M., 1999, Phys.Rev. D59, 103001

\bibitem[\protect\citeauthoryear{Spergel et al.}{2007}]{Spergel07}
Spergel D.N. et al., 2007, ApJS, 170, 377

\bibitem[\protect\citeauthoryear{Sunyaev \& Zeldovich}{1980}]{SZ}
 Sunyaev R.~A., Zeldovich I.~B., 1980, MNRAS, 190, 413 

\bibitem[\protect\citeauthoryear{Szapudi \& Szalay}{1999}]{Szapudi}
Szapudi I., Szalay A.S. Astrophys.J. 515 (1999) L43

\bibitem[\protect\citeauthoryear{Tegmark}{1997}]{teg} 
Tegmark M., Phys.Rev. D55 (1997) 5895-5907

\bibitem[\protect\citeauthoryear{Verde,\& Spergel}{2002}]{VerdeSperg02}
Verde L., Spergel D.N.,  2002, Phys. Rev. D65, 043007

\bibitem[\protect\citeauthoryear{Wang \& Kamionkowski}{2000}]{WangKam00}
Wang L., Kamionkowski M., 2001, Phys. Rev. D61, 3504

\bibitem[\protect\citeauthoryear{Yadav \& Wandelt}{2008}]{YaWa08}
Yadav A.~P.~S., Wandelt B.~D., 2008, PhRvL, 100, 181301

\bibitem[\protect\citeauthoryear{Yadav et al.}{2008}]{Yadav08} 
Yadav A.~P.~S., Komatsu E., Wandelt B.~D., Liguori M., Hansen F.~K., 
Matarrese S., 2008, ApJ, 678, 578 

\bibitem[\protect\citeauthoryear{Yadav, Komatsu \& Wandelt}{2007}]{YKW}
Yadav A.~P.~S., Komatsu E., Wandelt B.~D., 2007, ApJ, 664, 680 

\bibitem[\protect\citeauthoryear{Zahn \& Zaldarriaga}{2006}]{ZZ06} 
Zahn O., Zaldarriaga M., 2006, ApJ, 653, 922 

\end{thebibliography}
\end{document}